\documentclass[12pt]{article}
\usepackage{amsmath}
\usepackage{amsmath}
\usepackage{latexsym}
\usepackage{bbm}
\usepackage{amssymb}
\usepackage{lipsum}

\usepackage{tikz}
\usetikzlibrary{decorations.pathmorphing}
\usetikzlibrary{decorations.markings}

\usepackage{hyperref}
\usepackage[all]{hypcap}

\def \vol {\text{vol}\,}
\def \slc {\text{SL}(2, \mathbb{C})}

\def\d{\delta}

\def\l{\lambda}

\def\r{\rho}

\def\s{\sigma}

\def\and{{\rm and}}

\def\IC{\mathbbm C}

\newcommand{\be}{\begin{equation}}
\newcommand{\bea}{\begin{eqnarray}}

\newcommand{\ee}{\end{equation}}
\newcommand{\eea}{\end{eqnarray}}

\begin{document}
\vspace*{-1.0in}
\thispagestyle{empty}
\begin{flushright}
CALT-TH-2019-021\\
QMUL-PH-19-15
\end{flushright}

\normalsize
\baselineskip = 18pt
\parskip = 6pt

\vspace{1.0in}

{\Large \begin{center}
{\bf Unified Formalism for 6D Superamplitudes \\ Based on a Symplectic Grassmannian}
\end{center}}

\vskip 0.2in

\begin{center}
{\large John H. Schwarz$^1$, and Congkao Wen$^2$}

\vskip 0.15in

\textit{\small ${}^{1}$ Walter Burke Institute for Theoretical Physics,\\
California Institute of Technology,
Pasadena, CA 91125, USA.} \\  \vspace{6pt}

\textit{\small ${}^{2}$ Centre for Research in String Theory, School of Physics and Astronomy\\
Queen Mary University of London,
Mile End Road, London, E1 4NS,
United Kingdom}\\

\end{center}
\vspace{.25in}

\begin{center}
\textbf{Abstract}
\end{center}
\begin{quotation}

Recently, twistor-like formulations of tree amplitudes involving $n$ massless particles have been proposed for various 6D supersymmetric theories. The formulas are based on two different forms of the scattering equations:  one based on rational maps and the other based on polarized scattering equations. We show that both formulations can be interpreted in terms of a symplectic (or complex Lagrangian) Grassmannian, $\mathbb{LG}(n, 2n)$, and that they correspond to different ways of fixing the ${\rm GL}(n, \mathbb{C})$ symmetry of $\mathbb{LG}(n, 2n)$. This provides an understanding of the equivalence of these different-looking formulas, and it leads to new twistor-like formulas for 6D superamplitudes.

\end{quotation}

\vspace{1cm}

\thispagestyle{empty}

\vfill
\vskip 5.mm
\hrule width 5.cm
\vskip 2.mm
{
\noindent  {\scriptsize e-mails:  {\tt jhs@theory.caltech.edu, c.wen@qmul.ac.uk} }
}

\setcounter{footnote}{0}
\setcounter{page}{0}

\newpage

\tableofcontents

\newpage

\section{Introduction}

Recently, twistor-like formulations of tree amplitudes have been introduced for a wide range of interesting 6D supersymmetric theories based on rational maps~\cite{Heydeman:2017yww, Cachazo:2018hqa, Heydeman:2018dje}. Rational maps had been utilized previously for the study of superamplitudes of 4D and 3D
theories~\cite{Cachazo:2012da, Cachazo:2012kg, Cachazo:2012pz, Cachazo:2013iaa}. In particular, the 6D
formulas extend the well-known twistor formulation of the scattering amplitudes for 4D $\mathcal{N}=4$
super Yang--Mills (SYM) \cite{Witten:2003nn, Roiban:2004yf}. In the spirit of the CHY construction of
$n$-particle scattering amplitudes~\cite{Cachazo:2013hca}, the formulas can be schematically summarized in
the following form,
\bea
\mathcal{A}_n = \int d\mu^{\rm 6D}_n \, \mathcal{I}_L \, \mathcal{I}_R \, ,
\eea
where the measure $d\mu^{\rm 6D}_n$, which is theory-independent, encodes the general information of 6D
massless kinematics (such as momentum conservation $\sum_i p_i=0$). Explicitly, the measure is given in
(\ref{eq:rationalMap}) (for even $n$), and it can be viewed as a map from 6D kinematics to punctures of a Riemann sphere.
Due to certain peculiarities of the 6D spinor-helicity formalism, the construction of the measure based on
rational maps treats amplitudes with an even and an odd number of particles differently.

The factors $\mathcal{I}_L$ and $\mathcal{I}_R$ in the integrand form a left-right double copy that
contains the dynamical information for the specific theory under consideration. For instance,
(\ref{eq:M5-ILIR}) presents the formula describing scattering amplitudes of the world-volume theory of a
probe M5-brane in 11D Minkowski spacetime.  As we will review later, analogous formulas were proposed for
various other 6D supersymmetric theories including the world-volume theory of a probe D5-brane in 10D
Minkowski spacetime, 6D maximal SYM, and 6D $(2,2)$ and $(2,0)$ supergravity, as well as 4D ${\cal N}=4$ SYM on the Coulomb branch.

In recent work~\cite{Geyer:2018xgb}, Geyer and Mason introduced a different formulation of the scattering
amplitudes for the aforementioned 6D supersymmetric theories (except for $(2,0)$ supergravity). The new
formulas are based on ``polarized scattering equations", which generalize a 4D version of the scattering
equations of ambitwistor string theory that was proposed previously in~\cite{Geyer:2014fka}.\footnote{The polarized scattering equations
have also been extended to scattering amplitudes of 10D and 11D supersymmetric
theories~\cite{Geyer:2019ayz}.} As we commented previously, the formulas based on the rational maps describe
6D scattering amplitudes with $n$ even and $n$ odd differently. One of the nice properties of the
formulation based on polarized scattering equations is that the formulas treat the amplitudes with an even
and an odd number of particles in the same way.\footnote{In the special case of $n=3$ there is a degeneracy of the kinematics that must be addressed.}

In this paper we prove the equivalence of these two seemingly different constructions, as must be the case if they are both correct. The key observation is that the two formulas can be viewed as different ``gauge choices" of a larger geometric object: the symplectic (or complex Lagrangian) Grassmannian $\mathbb{LG}(n,2n)$, where $n$ is the number of scattering particles.\footnote{$\mathbb{LG}(n,2n)$ may be identified with the homogeneous space ${\rm USp}(2n)/{\rm U}(n)$, which has $n(n+1)/2$ complex dimensions. However, we will be led to a different realization of this space.} As already pointed out in~\cite{Cachazo:2018hqa}, the symplectic Grassmannian structure emerges naturally in the formulas based on rational maps. We will show
that this is also the case for the formulas based on the polarized scattering equations.

Recall that in general a Grassmannian $Gr(k,n)$ describes the space of  $k$-planes in $n$-dimensional space. It enjoys a ${\rm GL}(k, \IC)$ symmetry, the group of linear transformations that leaves the $k$-plane invariant, that should be divided out. We will show that the formulations based on rational maps and polarized scattering equations correspond to two different ways of fixing the ${\rm GL}(n, \IC)$ symmetry of the Grassmannian
$\mathbb{LG}(n,2n)$. This provides a unification of these two seemingly different constructions. ${\rm GL}(n, \IC)$ transformations of the integrands of rational-map formulas also lead to new twistor-like formulas for the superamplitudes of all of the 6D theories mentioned previously.

The rest of the paper is organized as follows. Section~\ref{sec:rational-polarized} reviews the 6D scattering equations for both formulations: one based on rational maps and the other based on polarized scattering equations.  Section~\ref{sec:equiv} shows that both formulations can be interpreted in terms of a symplectic Grassmannian $\mathbb{LG}(n,2n)$, and that they correspond to different ${\rm GL}(n, \IC)$ gauge choices of $\mathbb{LG}(n,2n)$, which establishes the equivalence of these two formulations.  In section~\ref{sec:Fer-measure}, we apply ${\rm GL}(n, \IC)$ transformations of the integrands to obtain new formulas for 6D superamplitudes. We conclude in section~\ref{sec:conclusion}.

\section{Rational maps and polarized scattering equations}  \label{sec:rational-polarized}

This section will briefly review the 6D rational-map constructions
of~\cite{Heydeman:2017yww, Cachazo:2018hqa, Heydeman:2018dje}, and the polarized scattering equations introduced in~\cite{Geyer:2018xgb}. We formulate scattering amplitudes of massless particles in 6D using the 6D spinor-helicity formalism~\cite{Cheung:2009dc}, which expresses the massless momentum as
\begin{align} \label{eq:6Dspinor-helicity}
p^{AB} = \langle \lambda^A \lambda^{B} \rangle = \frac{1}{2} \epsilon^{ABCD} [ \tilde{\lambda}_{C}
\tilde{\lambda}_{D} ] \, ,
\end{align}
where $A, B=1,2,3,4$ are spinor indices of Lorentz group ${\rm Spin}(1,5)$. Here we have used the short-hand notation
\begin{align}
 \langle \lambda^A \lambda^{B} \rangle := \lambda^A_a \lambda^{B}_b \epsilon^{ab} \, , \quad
 [ \tilde{\lambda}_{A} \tilde{\lambda}_{B} ]: =\tilde{\lambda}_{A, \hat{a} } \tilde{\lambda}_{B, \hat{b}}
 \epsilon^{\hat{a} \hat{b} } \, ,
\end{align}
where $a, b$ and  $\hat{a}, \hat{b}$ are little-group indices. For a 6D massless particle, the little group
is ${\rm Spin}(4)\sim {\rm SU}(2)_L \times {\rm SU}(2)_R$, so $a, b$ in the above equation are the indices
of ${\rm SU}(2)_L$ with $a, b=1, 2$,  and  $\hat{a}, \hat{b}=\hat{1}, \hat{2}$ refer to ${\rm SU}(2)_R$.

In addition to the use of the spinor-helicity formalism, the two approaches, described in the two following subsections, have some other features in common. Perhaps the most basic is that to each of the $n$ massless external particles in the scattering amplitude we assign a coordinate $\sigma_i$, $i=1,2,\dots,n$, that is sometimes referred to as a puncture of the Riemann sphere (represented as the complex plane plus a point at infinity). These coordinates, which are distinct, are defined up to an overall common
$\slc_\sigma$ M{\"o}bius-group transformation,
\bea
\sigma_i \rightarrow \frac{a\sigma_i + b}{c\sigma_i + d}, \qquad ad-bc = 1.
\eea
This allows the coordinates of three of the punctures to be given arbitrary distinct values.
The contribution to the integration measure $d\mu^{\rm 6D}_n$ is then
$\prod_{i=1}^n d \sigma_i$/vol($\slc_\sigma$), which is defined in a standard way.

A function of the $\s$ coordinates $F(\{ \sigma_i \} )$ is said to have weight $w$ if it
transforms under the M{\"o}bius group by the rule
\bea
F \left( \left\{ \frac{ a\sigma_i +b} {c\sigma_i +d }\right\} \right)
= \left[ \prod_{i=1}^n(c\sigma_i + d)\right]^w F(\{ \sigma_i \} ).
\eea
It will turn out that the measure $d\mu^{\rm 6D}_n$ transforms with $w=-4$. Therefore, for the amplitude to be well-defined, the rest of the integrand, $\mathcal{I}_L \, \mathcal{I}_R$, must have weight 4. In practice, each of the two factors, $\mathcal{I}_L$ and $\mathcal{I}_R$, always has weight 2.

Another common feature of the two approaches is that for supersymmetric theories we also introduce Grassmann variables. We will follow~\cite{Heydeman:2017yww, Cachazo:2018hqa, Heydeman:2018dje} for the construction of supersymmetry, where Grassmann variables $\eta^I_a$ (as well as $\tilde{\eta}^{\tilde I}_{\hat a}$ in the case of non-chiral theories) will be introduced, here the indices $I$ and $\tilde I$ label the supersymmetries. This enables concise description of complete on-shell supermultiplets.

In the remainder of this section, we describe the bosonic integration measure, in
forms appropriate to the rational-maps and the polarized scattering equations approaches. We will discuss their supersymmetric extensions in section \ref{sec:Fer-measure} when we consider specific theories.

\subsection{Rational maps and symplectic Grassmannians}
Let us consider first the construction based on rational maps. For even $n$, the $n$-particle measure is given by
\bea \label{eq:rationalMap}
\prod_{i=1}^n \delta(p_i^2)\int d\mu^{\rm 6D}_n =\int \frac{\prod_{i=1}^n d\sigma_i\, \prod_{k=0}^{m} d^8
\rho_k}{\vol( \slc_\sigma \times \slc_\rho)} \frac{1}{V_n^2}\prod_{i=1}^n  \delta^6 \left( p^{AB}_i -
\frac{\langle \rho^{A}(\sigma_i)\, \rho^{B}(\sigma_i) \rangle }{\prod_{j\neq i} \sigma_{ij}}\right)\, ,
\eea
where $n=2m+2$. The delta functions on the right-hand side imply $p_i^2=0$, so we define the integration measure $d\mu^{\rm 6D}_n$ with all of the $\delta(p_i^2)$ factors removed. We have also defined
\bea
\sigma_{ij} =\s_i - \s_j \quad {\rm and} \quad V_n = \prod_{i<j} \s_{ij} \, ,
\eea
where the latter is a Vandermonde determinant. The delta functions of the 6D rational-map scattering equations impose the masslessness conditions $p^2_i=0$, as we mentioned, and also the conservation of total momenta $\sum_{i=1}^n p_i^{AB} =0$. The maps in the delta functions are given by degree-$m$ polynomials,
\bea
\rho^{A}_a(\sigma) = \sum_{k=0}^{m} \rho^A_{a,k}\, \sigma^k \, ,
\eea
which are determined up to an overall $\slc_\rho$ transformation, which is a complexification of ${\rm SU}(2)_L$, and its volume also is divided out.

The $\slc_\sigma$ transformations of the coordinates $\rho^A_{a,k}$ are determined by requiring that the expressions inside the delta functions in (\ref{eq:rationalMap}) are invariant. Then one can show (with some effort) that $d\mu^{\rm 6D}_n$ has weight $-4$.

As shown in~\cite{Cachazo:2018hqa}, by introducing $n$ additional $2 \times 2$ matrices $(W_i)^b_a$, the rational-map scattering equations (\ref{eq:rationalMap}) can be recast in the ``linear" form\footnote{Here ``linear" is in comparison with (\ref{eq:rationalMap}), where the maps $\r^A_{a}(\s_i)$ enter the constraints
quadratically. }
\begin{align} \label{eq:linear-1}
\int \! d\mu_{n}^{\text{6D}}  = \int\! \frac{\prod_{i=1}^n\, d\sigma_i\,  d^{4}W_i \prod_{k=0}^{m} d^8
\rho_k}{\vol( \slc_\sigma \times \slc_\rho)}  \prod_{i=1}^n \d^8 \!\left ( \l^A_{i,a} - (W_i)^b_a \r^A_{b}
(\s_i) \right ) \d\!\left(|W_i| - {1 \over\prod_{j\neq i}\sigma_{ij} } \right) \, ,
\end{align}
where we use the short-hand notation: $|W_i|={\rm det}\, W_i$. In this formula
the ${\rm SU}(2)$ indices $a$ and $b$ of $(W_i)^b_a$ refer to different groups. Specifically, $b$ is contracted with the ${\rm SU}(2)$ index of the moduli $\rho^A_{k, b}$, and therefore it is a global little-group index, whereas $a$ is associated with the little group of the $i$-th particle.

We can now integrate out the moduli $\rho^A_{a,k}$ of the maps~\cite{Cachazo:2013zc}, which leaves an integral over only the $\s_i$'s and the $W_i$ matrices. Then (\ref{eq:linear-1}) reduces to
\begin{align} \label{eq:emerge-Gr}
\int \! d\mu_{n}^{\text{6D}} =
 \int\! \frac{\prod_{i=1}^n d\sigma_i\, d^{4}W_i  }{\vol( \slc_\sigma \times \slc_W)} \prod^m_{k=0}
 \d^{2\times 4} \left(\sum_{i=1}^n (W_i)^b_a \s^k_i \l^{A, a}_{i}\right)\prod_{i=1}^n \d\left(|W_i|
 - {1 \over\prod_{j\neq i}\sigma_{ij} }\right) \, ,
\end{align}
where the action of $\slc_\rho$ has become $\slc_W$, which is the symmetry acting on the global little-group index $b$. We can now show the emergence of the symplectic Grassmannian. $(W_i)^b_a \s_i^k$ can be viewed as an $n\times 2n$ matrix:
\begin{equation} \label{eq:C-matrix}
{C}_{k, b;i, a}=(W_i)^b_a\, \s_i^k\, ,
\end{equation}
where we group the exponent $k$ with the global $\slc$ index $b$ and the index $i$ with the $i$-th little-group $\slc$ index $a$.
The matrix $C$ formed in this way satisfies the identity
\begin{align} \label{eq:identity}
C\cdot \Omega \cdot C^T =0 \,,
\end{align}
where $\Omega$ is the symplectic metric
\begin{equation}
\Omega= \begin{pmatrix}
0 & \mathbb{I}_n \\
- \mathbb{I}_n  & 0
\end{pmatrix} \, ,
\end{equation}
where $\mathbb{I}_n$ is the $n \times n$ identity matrix. If $M$ is a symplectic matrix belonging to ${\rm USp}(2n)$, satisfying $M^T \cdot \Omega\cdot M = \Omega$, the identity (\ref{eq:identity}) is invariant under the symplectic transformation $C\to C\cdot M^T$. It is proved by using the delta-function constraints and the theorem\footnote{This theorem is easy to establish by showing that the residues of poles vanish (whenever $K$ is a non-negative integer) and that there are suitable asymptotic properties for $K < n-1$.}
\begin{equation}\label{eq:sigforms}
\sum_{i=1}^n \frac{\sigma_i^K}{\prod_{j\neq i} \s_{ij}} =0 \quad {\rm for }\quad  K=0,1,\ldots, n-2\, .
\end{equation}
The scattering-equation constraints can then be encoded as
\bea \label{eq:RMap-linear}
 \prod_{k=0}^{m} \delta^{2\times 4} \left(\sum_{i=1}^n {C}_{k, b;i, a} \, {\lambda}^{A a}_i \right) \, .
\eea

For amplitudes with an odd number of particles, $n=2m+1$, there are odd-$n$ versions of
(\ref{eq:rationalMap}) and (\ref{eq:RMap-linear})~\cite{Cachazo:2018hqa}. Here we will only present the formula in the symplectic Grassmannian form, which is more relevant for the following discussions. For odd $n$  the delta function product in (\ref{eq:RMap-linear}) is replaced by
\bea \label{eq:RMap-linear-2}
 \delta^{4}  \left(\sum_{i=1}^n {C}_{m, b;i, a} \xi^b \, {\lambda}^{A a}_i \right)  \prod_{k=0}^{m-1}
 \delta^{2\times 4} \left(\sum_{i=1}^n  {C}_{k, b;i, a} \, {\lambda}^{A a}_i  \right) \, ,
\eea
where additional integration variables $\xi^b$ have been introduced~\cite{Cachazo:2018hqa}. There is an additional ${\rm GL}(1, \IC)$ symmetry acting on $\xi^b$ as well as a ``T-shift" symmetry that acts on $C_{k, b; i, a}$ as
\begin{align}
C_{k, b; i, a} &\rightarrow C_{k, b; i, a} - \alpha\, \xi_b \, \xi^c C_{k+1, c; i, a}\,, \qquad k=0, 1,
\cdots, m-1 \, , \\
C_{m, b; i, a} &\rightarrow C_{m, b; i, a} \, ,
\end{align}
for an arbitrary parameter $\alpha$. In this way, the number of integration variables and the number of the
delta functions continue to match. $C_{k, b; i, a}$ is still given by (\ref{eq:C-matrix}). Now
${C}_{k, b;i, a}$ (with $k=0,1,\ldots, m-1$) together with ${C}_{m, b;i, a} \xi^b$ form an $n\times 2n$
matrix.

In general, for both even and odd $n$, we can encode the 6D scattering equations in the Grassmannian form
\bea \label{eq:CLambda}
\delta^{4\times n} (C \cdot \Omega \cdot \Lambda^A ) \, ,
\eea
with $C \cdot \Omega \cdot C^T=0$. We have introduced a $2n$-dimensional vector $\Lambda^A$, which is also
a Lorentz spinor, built out of the spinor-helicity coordinates $\lambda^{A}_{i, a}$,
\begin{equation}
\Lambda^A :=\{ \lambda^{A}_{1, 1}, \lambda^{A}_{2, 1}, \ldots, \lambda^{A}_{n, 1},  \lambda^{A}_{1, 2},
\lambda^{A}_{2, 2}, \ldots, \lambda^{A}_{n, 2}\}  \,  .
\end{equation}
Invariance under symplectic transformations requires that $\Lambda^A \rightarrow M \cdot \Lambda^A$.

Let us verify that $C$ parametrizes $\mathbb{LG}(n,2n)$, as has been claimed, see e.g.~\cite{Symplectic}. First we can check
the dimension, which is supposed to be $n(n+1)/2$. To begin, $C$ is an $n \times 2n$ complex matrix, which has $2n^2$ complex dimensions. However,
it can be multiplied on the left by an arbitrary ${\rm GL}(n, \IC)$ matrix, without changing the scattering equations or the constraint equations. The latter are given by the $n \times n$ antisymmetric matrix equation $C \cdot \Omega \cdot C^T =0$. Altogether, we are left with
$2n^2 - n^2 - n(n-1)/2 = n(n+1)/2$ complex dimensions. The remaining requirement is that
USp$(2n)$ should be a ``global" symmetry to agree with the homogeneous space USp$(2n)/$U$(n)$.
This has been demonstrated to be the case in the preceding paragraphs.

For non-chiral theories, such as 6D SYM and supergravity, the integrands require the use of additional
conjugate variables ($\widetilde{W}_i$, and $\tilde{\xi}$ if $n$ is
odd). They satisfy analogous rational-map constraints,
\bea \label{eq:CLambda-2}
\delta^{4\times n} (\widetilde{C} \cdot \widetilde\Omega \cdot \tilde{\Lambda}_{A} ) \, ,
\eea
where $\widetilde{C}$  and $\widetilde\Omega$ are defined in the same way as ${C}$ and $\Omega$, but with
$W_i \rightarrow \widetilde{W}_i$, $a \rightarrow \hat{a}$, and $\xi \rightarrow \tilde{\xi}$ (if $n$ is
odd), and the helicity spinor $\tilde{\Lambda}_{A}:=\{ \tilde{\lambda}_{1, A, \hat{1}},  \ldots,
\tilde{\lambda}_{n, A, \hat{1}}, \tilde{\lambda}_{1, A, \hat{2}}, \ldots,
\tilde{\lambda}_{n, A, \hat{2}}\}$. Since (\ref{eq:emerge-Gr}) is already the integration measure that encodes the 6D kinematics, the conjugate delta functions (\ref{eq:CLambda-2}) in fact do not explicitly appear in the construction of amplitudes, but they are used to determine $\widetilde{W}_i$ and $\tilde{\xi}$ if these conjugate variables are involved in the integrands, as we will see, which is the case for non-chiral theories.

\subsection{Polarized scattering equations and symplectic Grassmannians}

Let us now review the 6D polarized scattering equations
that were introduced in~\cite{Geyer:2018xgb}. In this approach associated to each external particle there
is a null momentum $p_i^{AB}$ (and hence spinor-helicity coordinates $\lambda^{Aa}_i$) {\em and}
arbitrary nonzero polarization spinors $\epsilon_{i, a}$. Explicitly, the corresponding 6D measure
takes the form
\begin{align} \label{eq:polarized-eq}
\int d \mu_n^{6D} = \int  {\prod_{i=1}^n d \s_i \, d^{2} v_i \, d^{2} u_i \over {\rm vol} ( \slc_\s
\times  \slc_u  ) }\prod_{i=1}^n \delta( v_{i, a} \, \epsilon_i^a -1) \, \delta^4 (v_{i,a}
\lambda^{A a}_i - u_{i,a}\lambda^{A a} (\s_i)  ) \, .
\end{align}
Instead of the polynomials $\rho_a^A(\s)$ that appear in the case of the rational-map approach, the polarized
scattering equation approach uses rational functions $\lambda^{A a} (\s)$ given by
\begin{align}
\lambda^{A a} (\s)  = \sum_{j=1}^n { u^a_{j} \epsilon_{j,b} \l^{Ab}_j \over \s - \s_j }\, .
\end{align}
Thus, the delta functions imply that $\langle v_i \epsilon_i \rangle =1$ for all $i$, and
the scattering equations become
\bea
\langle v_i\, \lambda_i^A \rangle = \langle u_i\, \lambda^A(\sigma_i)\rangle
= \sum_{j \neq i} \frac{\langle u_i u_j\rangle \langle \epsilon_j \lambda_j^A \rangle}{\sigma_{ij}}.
\eea
The symmetry $\slc_u$, which we mod out, is a complexification of a global ${\rm SU}(2)_L$ symmetry
acting simultaneously on all of the $u^a_i$'s. ($\epsilon_i$, $v_i$, and $\lambda_i^A$ are inert.)

The little-group spinors $\epsilon_i$ satisfy the constraints $\langle v_{i} \, \epsilon_i \rangle =1$, as
shown explicitly in (\ref{eq:polarized-eq}). Moreover, the measure is invariant under the group
$\otimes_{i=1}^n {\rm SU(2)}_i$, where the $i$-th factor rotates $\epsilon_i$, $v_i$, and $\lambda_i^A$
simultaneously ($u_i$ is inert). This allows us to make a convenient choice of little-group frames,
specified by the $\epsilon_{i,a}$'s such as $\epsilon_{i, a} = (0, 1)$ for all $i$. For this choice the
delta-function constraint $\langle v_i \, \epsilon_i \rangle =1$ is solved by $v_{i, a} = (1, v_i)$, these
delta functions can be eliminated, and the integration over $v_{i, a}$ reduces to a one-dimensional integral.
However, as we will see, there are advantages to exhibiting all of the symmetries.

Based on the polarized scattering equations, very concise formulas were written down in~\cite{Geyer:2018xgb}
for the supersymmetric amplitudes of various 6D theories that had been studied previously
in~\cite{Heydeman:2017yww, Cachazo:2018hqa}. Even though the formulas for scattering amplitudes must be
the same in both approaches, the formulas based on rational maps and polarized scattering equations have
quite different structures. In particular, one of nice properties of the 6D polarized scattering equations
approach is that it treats scattering amplitudes with $n$ even and $n$ odd in a uniform way.

It is important that the measure based on the polarized scattering equations, given by delta functions
in (\ref{eq:polarized-eq}), can be expressed in the Grassmannian form
\bea \label{eq:Symplectic-V}
\int d \mu_n^{6D} = \int  {\prod_{i=1}^n d \s_i \, d^{2} v_i \, d^{2} u_i
\over {\rm vol} ( \slc)_\s \times  \slc_u  ) }\prod_{i=1}^n
\delta( v_{i, a} \, \epsilon_i^a -1)\delta^{4} ( V \cdot \Omega \cdot  \Lambda^A) \, .
\eea
The explicit form of the $n\times 2n$ matrix $V$ that follows from (\ref{eq:polarized-eq}) is
\bea \label{eq:V-matrix}
V_{i; j, a}= \left\{
                \begin{array}{ll}
                  v_{i, a} \qquad \qquad \quad {\rm if} \qquad  i=j \\
       - { \langle u_i u_j \rangle \over \s_{ij} } \epsilon_{j, a}   \qquad   {\rm if} \qquad i \neq j ,
                \end{array}
              \right.
\eea
where the indices are grouped as $V_{i; j, a}$, for $i, j= 1, 2, \ldots, n$ and $a=1,2$, such that $V$ is
an $n\times 2n$ matrix.
Crucially, like the matrix $C$ in the rational-map approach, $V$ is also symplectic, transforming as
$V \rightarrow VM^T$, that satisfies the ${\rm USp} (2n)$-invariant constraint
\bea \label{eq:sympletic}
V \cdot \Omega \cdot V^{T} =0\, .
\eea
This can be verified using the conditions $\langle v_i \, \epsilon_i \rangle =1$.
Note that the ${\rm USp}(2n)$ symmetry exhibited here actually contains the subgroup $\otimes_{i=1}^n {\rm SU(2)}_i$, as described earlier. Note, also, that the equations $\langle v_i \, \epsilon_i \rangle =1$
are only invariant under the subgroup.
Again, for non-chiral theories conjugate variables must be introduced,
and they satisfy the conjugate version of the polarized scattering equations,
\bea \label{eq:conjugate-V}
\delta^{4\times n}( \widetilde{V} \cdot \widetilde{\Omega} \cdot  \tilde{\Lambda}_A) \, ,
\eea
where $\widetilde{V}$ is given in the same form as (\ref{eq:Symplectic-V}) and (\ref{eq:V-matrix}), with $v_{i, a} \rightarrow \tilde{v}_{i, \hat a}, u_{i, a} \rightarrow \tilde{u}_{i, \hat a}$ and $\epsilon_i^a \rightarrow \tilde{\epsilon}_i^{\dot a}$.

To summarize, both the rational-map equations and the polarized scattering equations can be recast in the
symplectic Grassmannian form
\bea \label{eq:S-Grassmannian}
\delta^{4\times n}(S \cdot \Omega \cdot  \Lambda^A)  \quad {\rm with} \quad S \cdot \Omega \cdot S^T =0 \, ,
\eea
where $S=C$ or $S=V$. As in the usual Grassmannian representation of scattering amplitudes,
(\ref{eq:S-Grassmannian}) provides a geometric picture for the kinematics~\cite{ArkaniHamed:2009dn}.
$ S \cdot \Omega \cdot  \Lambda^A=0$ implies that $S$ is orthogonal to $\Lambda^A$, and $S \cdot \Omega
\cdot S^T =0$ means that $S^{T}$ contains $\Lambda^A$. The null symplectic property of $S$ then implies
that $(\Lambda^A)^T \cdot \Omega \cdot \Lambda^B =0$, which is momentum conservation.

\section{Equivalence of rational maps and polarized scattering equations} \label{sec:equiv}

In the previous section, we showed that the 6D scattering equations, in both the rational-map and the polarized scattering equations approaches, have a natural interpretation in terms of a symplectic Grassmannian. Here we will show that the geometric Grassmannian picture allows us to establish the equivalence of these two formulations.  The basic idea is that they are given by different ${\rm GL}(n, \IC)$ gauge fixings of a symplectic Grassmannian $\mathbb{LG}(n, 2n)$.

\subsection{4D scattering equations}

Before discussing the 6D case, let us begin with the simpler 4D case as a warmup exercise~\footnote{A similar argument was given in~\cite{He:2016vfi} for the 4D case.}. In contrast to 6D, where there is a single formula for the $n$-particle amplitude, in 4D the $n$-particle amplitude decomposes into $n-3$ sectors labeled by $(d, \tilde d)$, where $d+ \tilde d = n-2$ and $d = 1, 2, \dots, n-3$. Then the scattering equations that
play the same role as (\ref{eq:emerge-Gr}) for the $(d, \tilde d)$ sector, or equivalently the $(n,k)$ sector, where $k = d+1$, are encoded in the measure
\begin{align} \label{eq:4D-emerg-Gr}
\int d\mu^{\rm 4D}_{n,k} = &  \int  {\prod_{i=1}^n d\s_i \, d t_i/t_i  \over \vol(\slc_\sigma \times {\rm GL}(1, \IC))} 
  \prod_{m=0}^{d} \delta^2( \sum_{i=1}^n t_i \sigma_i^m \lambda_i^{\alpha} )
\prod_{\tilde{m}=0}^{\tilde{d}} \delta^2( \sum_{i=1}^n \tilde{t}_i \sigma_i^{\tilde{m}}
\tilde{\lambda}_i^{\dot \alpha} )  \, ,
\end{align}
where $\tilde{t}_i = {t_i}^{-1}  {1\over \prod_{j \neq i} \sigma_{ij} }$. In the above formula, we have used the standard spinor-helicity formalism for a 4D massless momentum,
\bea
p^{\alpha \, \dot {\alpha}}_i = \lambda^{\alpha}_i \tilde{\lambda}^{\dot \alpha}_i \, .
\eea
The constraints can be viewed in a Grassmannian form~\cite{ArkaniHamed:2009dn},
\bea
C_{\rm 4D} \cdot \Lambda^{\alpha} =0 \, , \qquad C^{\perp}_{\rm 4D} \cdot \tilde{\Lambda}^{\dot \alpha}
= 0 \, ,
\eea
where $\Lambda^{\alpha} := \{\lambda^{\alpha}_1, \lambda^{\alpha}_2, \cdots, \lambda^{\alpha}_n\}$, and
similarly for $\tilde{\Lambda}^{\dot \alpha}$. $C_{\rm 4D}$ is a $k \times n$ matrix with entries
given by $t_i \sigma_i^m$, namely
\bea \label{eq:CGr}
C_{\rm 4D} = \begin{pmatrix}
t_1 & t_2 & \cdots & t_n \\
t_1 \sigma_1 & t_2 \sigma_2 & \cdots & t_n \sigma_n
\\
\vdots & \vdots & \cdots & \vdots &
\\
t_1 \sigma^{d}_1 & t_2 \sigma^{d}_2 & \cdots & t_n \sigma^{d}_n
\end{pmatrix} \, .
\eea
Similarly, $C^{\perp}_{\rm 4D}$ is an $(n-k) \times n$ matrix with entries given by $\tilde{t}_i \sigma_i^{\tilde{m}}$. Due to the constraints $t_i \tilde{t}_i = {\prod_{j \neq i}
\sigma^{-1}_{ij} }$
and equation (\ref{eq:sigforms}), we have
\bea
C_{\rm 4D} \cdot [C_{4 \rm D}^{\perp}]^T = 0.
\eea
Namely, $C_{\rm 4D}^{\perp}$ is the complement of $C_{\rm 4D}$.

There is a ${\rm GL}(k, \IC)$ symmetry associated with this realization of the Grassmannian $Gr(k, n)$.\footnote{$Gr(k, n)$ can be identified with the homogeneous space ${\rm U}(n)/{\rm U}(k) \times {\rm  U}(n-k)$, which has $k(n-k)$ complex dimensions.} The scattering amplitudes are independent of how the ${\rm GL}(k, \IC)$ symmetry transformation is chosen. One choice of interest is to utilize the
${\rm GL}(k, \IC)$ symmetry to transform $C_{\rm 4D}$ into the form
\bea \label{eq:idkk}
C^{\prime}_{\rm 4D} := C_{{\rm GL}(k)} \cdot C_{\rm 4D} = \begin{pmatrix}
1 & 0 & \cdots & 0 & c_{1 \, k+1} & \cdots & c_{1 \, n} \\
0 & 1 & \cdots & 0 & c_{2 \, k+1} & \cdots & c_{2 \, n}
\\
\vdots & \vdots & \cdots & \vdots & \vdots & \cdots  & \vdots
\\
0 & 0 & \cdots & 1 & c_{k \, k+1} & \cdots & c_{k \, n}
\end{pmatrix} \, ,
\eea
where $C_{{\rm GL}(k)}$ is a ${\rm GL}(k, \IC)$ matrix. Obviously, the ${\rm GL}(k, \IC)$ transformation that turns
$C_{\rm 4D}$ into $C^{\prime}_{\rm 4D}$ is given by
\bea
C_{{\rm GL}(k)} = \begin{pmatrix}
t_1 & t_2 & \cdots & t_k \\
t_1 \sigma_1 & t_2 \sigma_2 & \cdots & t_k \sigma_k
\\
\vdots & \vdots & \cdots & \vdots &
\\
t_1 \sigma^{d}_1 & t_2 \sigma^{d}_2 & \cdots & t_k \sigma^{d}_k
\end{pmatrix}^{-1} \, .
\eea
After the ${\rm GL}(k, \IC)$ transformation that produces $C^{\prime}_{\rm 4D}$, the first set of delta functions in (\ref{eq:4D-emerg-Gr}), which encodes the 4D scattering equations, reduces to
\bea \label{eq:4D-polarized}
\prod_{i=1}^k \delta^2(\lambda^{\alpha}_i - \sum_{J=k+1}^n c_{i\, J} \lambda^{\alpha}_J) \, ,
\eea
where
\bea
c_{i \, J}  = {t_J \prod_{j\neq i} \sigma_{jJ} \over t_i  \prod_{j\neq i} \sigma_{ji} } \, ,
\eea
with $i \in \{1, 2, \cdots, k\}$ and $J \in \{ k+1, \cdots, n \}$. Define
\bea
t'_J := t_J \prod_{j} \sigma_{jJ}  \, , \qquad  t_i' := {1\over t_i  \prod_{j\neq i} \sigma_{ji}} \, ,
\eea
$c_{i \, J}$ can be expressed as
\bea
c_{i \, J}  = {t_i'  t'_J \over \sigma_{i J}} \, .
\eea
Via a corresponding $GL(\tilde{k})$ transformation, we fix the complement $C_{\rm 4D}^{\perp}$ to be
\bea
(C_{\rm 4D}^{\perp})' = \begin{pmatrix}
- c_{1 \, k+1} & \cdots & -c_{k \, k+1} & 1 & 0 & \cdots & 0 \\
- c_{1 \, k+2} & \cdots & -c_{k \, k+2} & 0 & 1 & \cdots & 0
\\
\vdots & \cdots & \vdots & \vdots & \vdots & \cdots  & \vdots
\\
- c_{1 \, n} & \cdots & - c_{k \, n} & 0 & 0 & \cdots & 1
\end{pmatrix} \, .
\eea
With these results, the 4D delta functions in (\ref{eq:4D-emerg-Gr}) become
\bea \label{eq:4D-polarized-2}
\prod_{i=1}^k \delta^2(\lambda^{\alpha}_i - \sum_{J=k+1}^n { t_i' t'_J  \lambda^{\alpha}_J \over
\sigma_{i\, J}}  )
\prod_{J=k+1}^{n} \delta^2(\tilde{\lambda}^{\dot \alpha}_J - \sum_{i=1}^k { {t}'_i
 {t}_J'  \tilde{\lambda}^{\dot \alpha}_i \over \sigma_{i\, J}}  ) \, .
\eea
This encodes the 4D version of the polarized scattering equations first proposed
in~\cite{Geyer:2014fka}.

In conclusion, a class of 4D theories, including ${\cal N} =4$ SYM, have $n-3$ sectors.\footnote{In the case of DBI-like theories, such as the super D3-brane, ${\cal A}_{n,k}$  is nonzero only if $n$ is even and $k=n/2$. The latter means that only the helicity-preserving amplitude, ${\cal A}_{n, n/2}$, is non-zero.}
The amplitude of the $k$-th sector, ${\cal A}_{n,k}$, is described by $Gr(k, n)$ with $k=2, 3, \ldots, n-2$. Therefore, the $n$-particle amplitudes have a ${\rm U}(n)$ symmetry in each sector. The ${\rm U}(1)^n$ subgroup corresponds to the product of little groups of the $n$ particles.
Different choices of ``gauge fixing" of the ${\rm GL}(k)$ symmetry of the formulas given above leads to different forms of the scattering equations.
Let us now apply this approach to 6D, where we will find that there is a single sector based on the symplectic Grassmannian $\mathbb{LG}(n, 2n)$.

\subsection{6D Scattering equations}

We now consider the scattering equations in 6D beginning with the case of even $n$. Recall that the
symplectic Grassmannian $\mathbb{LG}(n, 2n)$ obtained from rational maps in (\ref{eq:C-matrix}) takes the form
\bea \label{eq:C-W-matrix}
C = \begin{pmatrix}
C_{1} & C_{2}
\end{pmatrix} \, ,
\eea
where $C_{a}$ are $n\times n$ matrices
\bea \label{eq:C-W-matrix-1}
C_{a} = \begin{pmatrix}
(W_1)^1_a &(W_2)^1_a &  \cdots & (W_n)^1_a  \\
(W_1)^2_a &(W_2)^2_a &  \cdots  &(W_n)^2_a  \\
(W_1)^1_a \sigma_1 & (W_2)^1_a \sigma_2 &  \cdots  &(W_n)^1_a\sigma_n
\\
(W_1)^2_a \sigma_1 & (W_2)^2_a \sigma_2 &  \cdots  &(W_n)^2_a\sigma_n
\\
\vdots &\vdots &  \cdots & \vdots
\\
(W_1)^1_a \sigma^{{n\over 2}-1}_1   &(W_2)^1_a \sigma^{{n\over 2}-1}_2   & \cdots  &(W_n)^1_a
\sigma^{{n\over 2}-1}_n
\\
(W_1)^2_a \sigma^{{n\over 2}-1}_1  & (W_2)^2_a \sigma^{{n\over 2}-1}_2  & \cdots  &(W_n)^2_a
\sigma^{{n\over 2}-1}_n
\end{pmatrix} \, ,
\eea
and $a=1,2$ is the little-group index.

Let us compare this with the $V$ matrix of the polarized scattering equations. In the convenient choice we
made for $\epsilon_i^a$ and $v_i^a$, namely $\epsilon_i^a = (0,1)$ and $v_i^a = (1, v_i)$, the explicit
form of $\mathbb{LG}(n, 2n)$ in (\ref{eq:V-matrix}) is
\bea \label{eq:V-matrix-2}
V =  \begin{pmatrix}
\mathbb{I}_n & V_{2}
\end{pmatrix} \, ,
\eea
where $V_2$ is an $n\times n$ symmetric matrix
\bea \label{eq:V-matrix-3}
V_{2} =  \begin{pmatrix}
 v_1  & -{\langle u_1 u_2 \rangle \over \sigma_{12}} & \cdots &  -{\langle u_1 u_n \rangle \over
 \sigma_{1n}} \\
 -{\langle u_2 u_1 \rangle \over \sigma_{21}} & v_2 & \cdots & -{\langle u_2 u_n \rangle \over
 \sigma_{2n}}
\\
  \vdots &  \vdots &  \cdots & \vdots
  \\
-{\langle u_{n-1} u_1 \rangle \over \sigma_{n-1 \, 1}}  &  -{\langle u_{n-1} u_2 \rangle \over \sigma_{n-1 \, 2}} &  \cdots &  -{\langle u_{n-1} u_{n}
\rangle \over \sigma_{n-1 \, n}}
\\
  -{\langle u_n u_1 \rangle \over \sigma_{n1}} & -{\langle u_n u_2 \rangle \over \sigma_{n2}} & \cdots &
  v_n
\end{pmatrix} \, .
\eea

Just as in the case of 4D, to bring the $C$ matrix in (\ref{eq:C-W-matrix}) into the form of
the $V$ matrix in (\ref{eq:V-matrix-2}), we need to perform a ${\rm GL}(n, \IC)$ transformation. The ${\rm GL}(n, \IC)$ matrix
\begin{align} \label{eq:GL(n)}
C_{{\rm GL}(n)} = (C_{1})^{-1} \,
\end{align}
does the job. Then the ${\rm GL}(n, \IC)$ transformed matrix $C^{\prime}$ is
\bea
C^{\prime} = C_{{\rm GL}(n)} \cdot C = \begin{pmatrix} \mathbb{I}_n & (C_{1})^{-1} \cdot C_{2} \end{pmatrix}\, .
\eea
The constraint $C \cdot \Omega  \cdot  C^T = C_1  \cdot  C_2^T -C_2 \cdot  C_1^T =0$
implies that the $n\times n$ sub-matrix $(C_{1})^{-1} \cdot C_{2}$ is symmetric. This fact allows us to map each entry of $C^{\prime}$ to that of the $V$
matrix in (\ref{eq:V-matrix-2}) and to make the identification
\bea
(C_{1})^{-1} \cdot C_{2} = V_{2} \, .
\eea

For the entries on the diagonal part of the sub-matrix $V_{2}$, we identify
\bea \label{eq:vi}
v_i =  {|C_{[i]}| \, |C_{{\rm GL}(n)}| } \, ,
\eea
where $C_{[i]}$ is an $n \times n$ matrix of the form
\bea \label{eq:C-W-matrix-2}
C_{[i]} &=& \begin{pmatrix}
(W_1)^1_1 & \cdots & (W_i)^1_2 & \cdots & (W_n)^1_1 \\
(W_1)^2_1 & \cdots & (W_i)^2_2 & \cdots & (W_n)^2_1 \\
(W_1)^1_1 \sigma_1 & \cdots & (W_i)^1_2 \sigma_i & \cdots & (W_n)^1_1 \sigma_n
\\
(W_1)^2_1 \sigma_1 & \cdots & (W_i)^2_2 \sigma_i & \cdots & (W_n)^2_1 \sigma_n
\\
\vdots & \cdots & \vdots & \cdots & \vdots
\\
(W_1)^1_1 \sigma^{{n\over 2}-1}_1 & \cdots &  (W_i)^1_2 \sigma^{{n\over 2}-1}_i & \cdots & (W_n)^1_1
\sigma^{{n\over 2}-1}_n
\\
(W_1)^2_1 \sigma^{{n\over 2}-1}_1 & \cdots &  (W_i)^2_2 \sigma^{{n\over 2}-1}_i & \cdots & (W_n)^2_1
\sigma^{{n\over 2}-1}_n
\end{pmatrix} \, .
\eea
So $C_{[i]}$ only differs from $C_{1}$ given in (\ref{eq:C-W-matrix-1}) by a flip of one of the ${\rm SU}(2)$ indices of $W_i$ in the $i$-th column.
For the off-diagonal entries in $V_2$, we have
\bea \label{eq:uij}
{ \langle u_i u_j \rangle \over \s_{ij} } =  { (-1)^{i+j} \over \s_{ij} } {|C_{[i, j]}| \, |C_{{\rm GL}(n)}| } \, ,
\eea
where $C_{[i, j]}$ is a $(n{-}2)\times (n{-}2)$ matrix, given by
\begin{align} \label{eq:Cij}
C_{[i, j]} =
\begin{pmatrix}
\cdots & (W_{i-1})^1_1 &(W_{i+1})^1_1 & \cdots & (W_{j-1})^1_1 &(W_{j+1})^1_1 & \cdots  \\
 \cdots & (W_{i-1})^2_1 &(W_{i+1})^2_1 & \cdots & (W_{j-1})^2_1 &(W_{j+1})^2_1 & \cdots   \\
 \cdots & (W_{i-1})^1_1  \sigma_{i-1} &(W_{i+1})^1_1  \sigma_{i+1}& \cdots & (W_{j-1})^1_1
 \sigma_{j-1}&(W_{j+1})^1_1  \sigma_{j+1}& \cdots  \\
\cdots & (W_{i-1})^2_1  \sigma_{i-1} &(W_{i+1})^2_1  \sigma_{i+1} & \cdots & (W_{j-1})^2_1
\sigma_{j-1} &(W_{j+1})^2_1  \sigma_{j+1}& \cdots   \\
\cdots & \vdots & \vdots & \cdots & \vdots & \vdots & \cdots
\\
 \cdots & (W_{i-1})^1_1  \sigma^{{n\over 2} -2}_{i-1} &(W_{i+1})^1_1  \sigma^{{n\over 2} -2}_{i+1}
 & \cdots & (W_{j-1})^1_1  \sigma^{{n\over 2} -2}_{j-1} &(W_{j+1})^1_1  \sigma^{{n\over 2} -2}_{j+1}
 & \cdots   \\
 \cdots & (W_{i-1})^2_1  \sigma^{{n\over 2} -2}_{i-1} &(W_{i+1})^2_1  \sigma^{{n\over 2} -2}_{i+1}
 & \cdots & (W_{j-1})^2_1  \sigma^{{n\over 2} -2}_{j-1} &(W_{j+1})^2_1  \sigma^{{n\over 2} -2}_{j+1}
 & \cdots
\end{pmatrix} \, .
\end{align}
Namely, it is defined by removing the $i$-th and $j$-th columns as well as the last two rows (the rows with $\sigma^{n/2-1}$) of the matrix $C_1$ in (\ref{eq:C-W-matrix-1}). We note that $\langle u_i u_j \rangle$ in (\ref{eq:uij}) can be viewed as Pl${\rm \ddot{u}}$cker coordinates of a $Gr(2, n)$, whereas $|C_{[i, j]}|$ are Pl${\rm \ddot{u}}$cker coordinates of a $Gr(n-2, n)$. The Grassmannian duality between $Gr(2, n)$ and $Gr(n-2, n)$ then ensures that a solution exists for (\ref{eq:uij}).\footnote{We thank Alfredo Guevara for this observation.}

The ${\rm GL}(n, \IC)$ transformation procedure works in a similar fashion for the case of odd $n$. From (\ref{eq:RMap-linear-2}) we see that for odd $n$ the $C_a$ matrix is given by
\bea \label{eq:C-W-matrix-odd}
C_a = \begin{pmatrix}
(W_1)^b_a \xi_b \sigma^{{n-1 \over 2}}_1 & (W_2)^b_a \xi_b \sigma^{{n-1 \over 2}}_2 &  \cdots & (W_n)^b_a
\xi_b \sigma^{{n-1\over 2}}_n
\\
(W_1)^1_a &(W_2)^1_a & \cdots & (W_n)^1_a  \\
(W_1)^2_a & (W_2)^2_a & \cdots & (W_n)^2_a  \\
\vdots &\vdots & \cdots & \vdots
\\
(W_1)^1_a \sigma^{{n-3 \over 2}}_1 &(W_2)^1_a \sigma^{{n-3 \over 2}}_2 &  \cdots & (W_n)^1_a
\sigma^{{n-3\over 2}}_n
 \\
(W_1)^2_a \sigma^{{n-3 \over 2}}_1 &(W_2)^2_a \sigma^{{n-3 \over 2}}_2 &  \cdots & (W_n)^2_a
\sigma^{{n-3\over 2}}_n
\end{pmatrix} \, ,
\eea
and $C = (C_1 \,\,\, C_2)$. Again, there exists a ${\rm GL}(n, \IC)$ transformation that turns the $C$ matrix of the rational-map description into the $V$ matrix of the polarized scattering equations description, namely,
\bea \label{eq:GLn-general}
C_{{\rm GL}(n)} \cdot C = V \, ,
\eea
with $C_{{\rm GL}(n)}= (C_1)^{-1}$ and $V = ( \mathbb{I}_n \,\, \, V_2)$. The entries of $V_2$ matrix in terms of rational-map variables take the same forms as (\ref{eq:vi}) and (\ref{eq:uij}), and now $C_{[i]}, C_{[i, j]}$ are given by
\bea \label{eq:C-W-matrix-2-odd}
C_{[i]} &=& \begin{pmatrix}
(W_1)^b_1 \xi_b \sigma^{{n-1 \over 2}}_1  &  \cdots & (W_i)^b_1 \xi_b \sigma^{{n-1 \over 2}}_i & \cdots &
(W_n)^b_1 \xi_b \sigma^{{n-1\over 2}}_n
\\
(W_1)^1_1 & \cdots & (W_i)^1_2 & \cdots & (W_n)^1_1 \\
(W_1)^2_1 & \cdots & (W_i)^2_2 & \cdots & (W_n)^2_1 \\
\vdots & \cdots & \vdots & \cdots & \vdots
\\
(W_1)^1_1 \sigma^{n-3\over 2}_1 & \cdots &  (W_i)^1_2 \sigma^{n-3\over 2}_i & \cdots & (W_n)^1_1
\sigma^{n-3\over 2}_n
\\
(W_1)^2_1 \sigma^{n-3\over 2}_1 & \cdots &  (W_i)^2_2 \sigma^{n-3\over 2}_i & \cdots & (W_n)^2_1
\sigma^{n-3\over 2}_n
\end{pmatrix} \, ,
\eea
and
\begin{align} \label{eq:Cij-odd}
C_{[i, j]} =
\begin{pmatrix}
\cdots  & (W_{i-1})^b_1 \xi_b \sigma^{{n-3 \over 2}}_{i-1}   & (W_{i+1})^b_1 \xi_b
\sigma^{{n-3 \over 2}}_{i+1} & \cdots & (W_{j-1})^b_1 \xi_b \sigma^{{n-3 \over 2}}_{j-1} & (W_{j+1})^b_1
\xi_b \sigma^{{n-3 \over 2}}_{j+1} & \cdots \\
\cdots & (W_{i-1})^1_1 &(W_{i+1})^1_1 & \cdots & (W_{j-1})^1_1 &(W_{j+1})^1_1 & \cdots  \\
 \cdots & (W_{i-1})^2_1 &(W_{i+1})^2_1 & \cdots & (W_{j-1})^2_1 &(W_{j+1})^2_1 & \cdots   \\
\cdots & \vdots & \vdots & \cdots & \vdots & \vdots & \cdots
\\
 \cdots & (W_{i-1})^1_1  \sigma^{n-5 \over 2}_{i-1} &(W_{i+1})^1_1  \sigma^{n-5 \over 2}_{i+1} & \cdots &
 (W_{j-1})^1_1  \sigma^{n-5 \over 2}_{j-1} &(W_{j+1})^1_1  \sigma^{n-5 \over 2}_{j+1} & \cdots   \\
 \cdots & (W_{i-1})^2_1  \sigma^{n-5 \over 2}_{i-1} &(W_{i+1})^2_1  \sigma^{n-5 \over 2}_{i+1} & \cdots &
 (W_{j-1})^2_1  \sigma^{n-5 \over 2}_{j-1} &(W_{j+1})^2_1  \sigma^{n-5 \over 2}_{j+1} & \cdots
\end{pmatrix} \, .
\end{align}

Finally, we remark that the Jacobian arising from the ${\rm GL}(n, \IC)$ transformation of the integration measure is one. In particular, both measures of the rational maps and the polarized scattering equations are equivalent to that of the CHY scattering equations~\cite{Cachazo:2013iaa}.  Therefore
\begin{align} \label{eq:C-V}
\int d\mu^{\rm 6D}_n &= \int {d^n \s d^{3n} W \over \vol( \slc_{\sigma} \times \slc_{W} )}
\delta^{4\times n} (C \cdot \Omega \cdot \Lambda^A )
\\ \nonumber
&= \int {d^n \s d^n v d^{2n} u \over \vol( \slc_{\sigma} \times \slc_{u} )} \delta^{4 \times n} (V \cdot
\Omega \cdot \Lambda^A )\, .
\end{align}
This can be seen by viewing the $4n$-dimensional integral as an integration over the entries of $n\times 4$ sub-matrix of the $n \times 2n$  matrix of  the symplectic Grassmannian, the rest of entries are not independent once we solve the symplectic constraints using rational maps or polarized scattering equations.\footnote{The symplectic constraints $C\cdot \Omega \cdot C^T=0$ impose only  $n(n-1)/2$ conditions, which in general would not reduce the integral to $4n$ dimensional. This is the usual story of Grassmannian formulation of scattering amplitudes~\cite{ArkaniHamed:2009dn}, where one requires to choose appropriate contours to fix all the integration variables.} A ${\rm GL}(n, \IC)$ transformation generates a Jacobian $|C_{{\rm GL}(n)}|^4$ for the integration variables, which cancels that from the delta functions. In the above expression, we have used the fact that the determinant of each $W_i$ is constrained. Therefore it only gives rise to three integration variables and, if
$n$ is odd, it should be understood that it is necessary to include an additional integration over $\xi^a$ as well as an additional scaling symmetry and T-shift symmetry. In the second line of the equation we have set $\epsilon_{i, a}= (0, 1)$ and $v_{i, a} = (1, v_i)$.

Finally,  a conjugate version of the ${\rm GL}(n, \IC)$ transformation is also needed for non-chiral theories, namely
\bea \label{eq:GLn-general-2}
\widetilde{C}_{{\rm GL}(n)} \cdot \widetilde{C} = \widetilde{V} \, .
\eea
As we discussed early, such delta functions involving conjugate variables and $\tilde{\Lambda}_{A}$ do not appear explicitly in the formulas of scattering amplitudes, but they are used to determine conjugate variables in the integrands in terms of external kinematics $\tilde{\Lambda}_{A}$.

\section{GL(n,C) transformations and superamplitudes}  \label{sec:Fer-measure}

Scattering amplitudes for supersymmetric theories can be realized by introducing Grassmann variables $\eta^I_a$ with $I=1, \ldots, \mathcal{N}$ for a 6D chiral $(\mathcal{N}, 0)$ supersymmetric theory,
whereas for a non-chiral $(\mathcal{N}, \widetilde{\mathcal{N}})$ supersymmetric theory, we will require
both $\eta^I_a$ and $\tilde{\eta}^{\tilde I}_{\hat a}$ with $I=1, \ldots, \mathcal{N}$ and ${\tilde I}=1,
\ldots,  \widetilde{\mathcal{N}}$. With this setup, the supercharges for the $i$-th particle in the
amplitude of a chiral theory are given by
\bea
q_i^{A, I} = \lambda^A_{i , a} \eta^{a, I}_i \, , \quad \bar{q}_{i, I}^{A} = \lambda^A_{i , a} {\partial
\over \partial \eta^{I}_{a,i } }\,  ,
\eea
and for a non-chiral theory we also have
\bea
q_{i, A}^{\tilde I} = \tilde{\lambda}^{\hat{a}}_{i , A} \tilde{\eta}^{\tilde I}_{\hat a, i} \, , \quad
\bar{q}_{i, A, \tilde I} = \tilde{\lambda}^{\hat{a}}_{i , A}  {\partial \over \partial
\tilde{\eta}^{\hat a, \tilde I}_{ i} } \, .
\eea
Supersymmetry then implies conservation of the supercharges, which means that the amplitudes should be
annihilated by
\bea \label{eq:Q-charge}
Q^{A, I}  = \sum_{i=1}^n q_i^{A, I} \, , \qquad  \bar{Q}^{A, I}  = \sum_{i=1}^n \bar{q}_i^{A, I}  \, ,
\eea
as well as by $\tilde{Q}^{I}_A$ and  $\bar{\tilde{Q}}^{\tilde I}_A$ for a non-chiral theory.

The construction of supersymmetry presented here follows
closely~\cite{Heydeman:2017yww, Cachazo:2018hqa, Heydeman:2018dje}, and is different from those utilized
in~\cite{Geyer:2018xgb}.
As shown in~\cite{Cachazo:2018hqa}, for a chiral supersymmetric theory the fermionic measure that implements
supersymmetry contains
\bea \label{eq:Gr-fermion}
\prod_{I=1}^{\mathcal{N}} \delta^n (C \cdot \Omega \cdot \eta^I ) \, ,
\eea
where $\eta^I:=\{\eta^{I}_{1,1},\eta^{I}_{2,1}, \ldots, \eta^{I}_{n,1}, \eta^{ I}_{1,2},\eta^{I}_{2,2},
\ldots, \eta^{I}_{n,2}\}$.  For a non-chiral theory this becomes
\bea \label{eq:Gr-fermion-2}
\prod_{I=1}^{\mathcal{N}} \delta^n (C \cdot \Omega \cdot \eta^I )
\prod_{{\tilde I} =1}^{\widetilde {\mathcal{N}}} \delta^n (\widetilde{C} \cdot \widetilde{\Omega} \cdot
\tilde{\eta}^{\tilde I} ) \, .
\eea
The fermionic measures take the same form as bosonic ones in (\ref{eq:CLambda}) with $\Lambda^A$ and
$\tilde\Lambda_A$ replaced by the Grassmann variables $\eta^I$ and $\tilde{\eta}^{\tilde I}$. Therefore, just as the
bosonic delta functions imply momentum conservation, the fermionic ones imply conservation of the
supercharges. Only the conservation of a mutually anticommuting set of supercharges (namely, $Q^{A, I}, \tilde{Q}^{I}_A$ given in (\ref{eq:Q-charge})) can be realized by fermionic delta function factors in the
amplitudes. Conservation of the remaining supercharges can be shown by acting the differential operators
of $\bar{Q}^{A, I}, \bar{\tilde{Q}}^{\tilde I}_A$ on the fermionic delta functions, which simply replaces
$\delta^n (C \cdot \Omega \cdot \eta^I )$ by $\delta^n (C \cdot \Omega \cdot \Lambda^A)$, and the later is satisfied due to the scattering equations.

\subsection{GL(n,C)}

The ${\rm GL}(n, \IC)$ transformation that relates the scattering equations appearing in the rational-map and the polarized scattering equation approaches also acts on the fermionic measures, which leads to fermionic counterparts of the polarized scattering equations
\begin{align} \label{eq:C-V-fermion}
\delta^n (V \cdot \Omega \cdot \eta^I )  = |C_{{\rm GL}(n)}| \, \delta^n(C \cdot \Omega \cdot \eta^I) \, , \qquad
\delta^n (\widetilde{V} \cdot \widetilde{\Omega} \cdot \tilde{\eta}^{\tilde{I}} )
= |\widetilde{C}_{{\rm GL}(n)}| \, \delta^n( \widetilde{C} \cdot \widetilde{\Omega} \cdot
\tilde{\eta}^{\tilde{I}} ) \, ,
\end{align}
for $I=1, 2, \ldots, {\cal N}$ and $\tilde{I}=1, 2, \ldots, \widetilde{\cal N}$, so these delta functions have multiplicity $n {\cal N}$ and $n \widetilde{\mathcal{N}}$, respectively. Here $|C_{{\rm GL}(n)}|$ and $|\widetilde{C}_{{\rm GL}(n)}|$ are the Jacobians of the ${\rm GL}(n, \IC)$ transformations.

Both sides of (\ref{eq:C-V-fermion}) are degree-$n$ polynomials of $\eta_{i,a}^I$, from which we can straightforwardly deduce the Jacobian $|C_{{\rm GL}(n)}|$. Consider a particular term of the polynomial, for instance,
\bea
\prod_{i\in Y}  \eta_{i, a}^I \eta_{i}^{I, a}  \, ,
\eea
where  $Y$ is an arbitrary length-$(n/2)$ subset of the particle labels, $\{1,2,\ldots, n\}$, and
$\overline{Y}$ is defined to be the complement of $Y$. Using $|W_i|= \prod_{j\neq i} \sigma_{ij}^{-1}$, we find that the determinant of the transformation $ {\rm GL}(n, \IC)$ can be expressed as
\bea \label{eq:Jaco-GLn}
{ |C_{{\rm GL}(n)}| }= { |U_{Y, \overline{Y}}|\, (V_n)^{1\over 2} \over |X_{Y, \overline{Y}}|^{1\over 2} } \, ,
\eea
which will become useful later. Again $V_n$ is the Vandermonde determinant, and $U_{Y,\overline{Y}}$ and $X_{Y, \overline{Y}}$ are $n/2 \times n/2$ matrices,
\bea \label{eq:U-matrix}
(U_{Y, \overline{Y}})_{iJ} = {\langle u_i u_J\rangle \over \sigma_{iJ}} \, , \qquad
(X_{Y, \overline{Y}})_{iJ} = {1 \over \sigma_{iJ}} \, ,
\eea
for $i \in Y$ and $J \in \overline{Y}$. The fact that the RHS of (\ref{eq:Jaco-GLn}) is independent of the
choice of $Y$ and ${\bar Y}$ can be made manifest using the identity~\cite{Roehrig:2017wvh}
\bea \label{eq:X-id}
 {|U_{Y, \overline{Y}}|^2 \over |X_{Y, \overline{Y}}|}= {\rm Pf}\, U_n\, ,
\eea
where $U_n$ is an $n\times n$ matrix with entries
\bea \label{eq:Un}
(U_n)_{ij} = \frac{\langle u_i u_j \rangle^2}{\sigma_{ij}} \, ,
\eea
for $i, j= 1,2, \ldots, n$. Combining (\ref{eq:Jaco-GLn}) and (\ref{eq:X-id}) gives
\bea
|C_{{\rm GL}(n)}|^2 = V_n \, {\rm Pf}\, U_n.
\eea
Since ${\rm Pf}\, U_n$ has conformal weight $-1$ and $V_n$ has weight $1-n$, it follows
that $|C_{{\rm GL}(n)}|$ has weight $- n/2$. Note that the weights balance in (\ref{eq:C-V-fermion})
where $\delta^n(V \cdot \Omega \cdot \eta^I)$ has weight $0$ and $\delta^n(C \cdot \Omega \cdot \eta^I)$
has weight $n/2$ for each $I$. There are analogous relations with tildes.

For odd $n$, we can compare the coefficients of a term
\bea
\prod_{i\in Y}  \eta_{i, a}^I \eta_{i}^{I, a} \, \eta_{k}^{I, 1} \, ,
\eea
where $Y$ is a length-$(n{-}1)/2$ sub-set of $\{1, 2, \ldots, k{-}1, k{+}1, \ldots, n \}$, namely the label $k$ is removed. From this consideration, we find that the determinant of the ${\rm GL}(n, \IC)$ transformation can be expressed as
\bea \label{eq:CGL-odd}
{|C_{{\rm GL}(n)}|^2}=  {1 \over ((W_k)^b_1 \, \xi_b)^2}  \, V_{n-1}^{(k)}\,   {\rm Pf} \,U^{(k)}_{n-1} \, .
\eea
Here $U^{(k)}_{n-1}$ is a $(n{-}1) \times (n{-}1)$ matrix that is defined in the same way as (\ref{eq:Un}), the labels run over $(n{-}1)$ particles, $\{1, 2, \ldots, k{-}1, k{+}1, \ldots, n \}$, and $V^{(k)}_{n-1}=\prod_{i<j} \sigma_{ij}$
for $i, j \in \{1, 2, \ldots, k{-}1, k{+}1, \ldots, n \}$. The result is independent of the choice of $k$.

With the above results, we are now ready to apply ${\rm GL}(n, \IC)$ transformations to the tree-level amplitude formulas based on rational maps presented in~\cite{Heydeman:2017yww, Cachazo:2018hqa, Heydeman:2018dje}, which will lead to new formulas for all of these superamplitudes.

\subsection{M5 and D5-brane}

We begin with the world-volume theory of a single probe M5-brane. It is a Born--Infeld-like theory that
only has nonzero amplitudes for $n$ even. It has $(2,0)$ supersymmetry, so $I=1,2$, and its on-shell
spectrum can be packaged into an on-shell ``superfield" or ``super wave function"
\bea \label{eq:M5-field}
\Phi(\eta) = \phi + \eta^a_I \psi_a^I + \eta^a_I  \eta^{I, b}  B_{ab} + \ldots + (\eta)^4 \bar{\phi} \, ,
\eea
where $B_{ab} = B_{ba}$ encodes the on-shell modes of the self-dual two-form. The formula that describes
the tree amplitudes of this theory is given by~\cite{Heydeman:2017yww}
\bea \label{eq:M5-formula}
\mathcal{A}^{\rm M5}_{n}
=
 \int d \mu_n^{\rm 6D}  \, {\cal I}_L^{(2,0)} \,  {\cal I}_R^{\rm DBI} \, ,
\eea
where the factors ${\cal I}_L^{(2,0)}$ and ${\cal I}_R^{\rm DBI}$ in the integrand are
\bea \label{eq:M5-ILIR}
{\cal I}_L^{(2,0)}
= \delta^{2\times n} (C \cdot \Omega \cdot \eta^I ) \, V_n\,  { {\rm Pf}^{\prime} S_n  } \, , \qquad
{\cal I}_R^{\rm DBI}
=  \left( {\rm Pf}^{\prime} S_n \right)^2 =  {\rm det}^{\prime} S_n\, .
\eea
Recall that $V_n$ is the Vandermonde determinant, and the bosonic measure is given by
\bea \label{eq:RM-measure}
\int d \mu_n^{\rm 6D}  = \int {d^n \s d^{3n} W \over \vol( \slc_{\sigma} \times \slc_{W} )} \delta^{4\times n}
(C \cdot \Omega \cdot \Lambda^A )\, .
\eea
The matrix $S_n$ entering the integrand is an $n\times n$ matrix. It is antisymmetric and has rank $n{-}2$,
with entries given by
\bea \label{eq:sij}
[S_n]_{ij} = { p_i \cdot p_j \over \sigma_{ij}}\, .
\eea
The reduced Pfaffian of  $S_n$ is defined as
\bea \label{eq:Sn-matrix}
{\rm Pf}^{\prime} S_n = {(-1)^{k+l} \over \sigma_{kl}} {\rm Pf} (S_n)^{kl}_{kl} \, ,
\eea
where $(S_n)^{kl}_{kl}$ is an $(n{-}2) \times (n{-}2)$ matrix with the $k$-th and $l$-th rows and columns of $S_n$ removed, and the result is independent of the choice of $k, l$. Since ${\rm Pf}' S_n$ has conformal weight $w=1$, we see that ${\cal I}_L^{(2,0)}$ and ${\cal I}_R^{\rm DBI}$
each have conformal weight $w=2$, as required. The factor $ {\cal I}_R^{\rm DBI}$, which is only well-defined
for even $n$, appears in all DBI-type theories. These theories only have nonvanishing amplitudes when $n$ is even.

Applying the result in (\ref{eq:C-V-fermion}), the ${\rm GL}(n, \IC)$ transformation leads to a new representation of the superamplitudes,
\bea
\mathcal{A}^{\rm M5}_{n}
=
 \int d\mu_n^{\rm 6D}  \, \delta^{2 \times n} (V \cdot \Omega \cdot \eta^I ) \,  {V_n
 \over |C_{{\rm GL}(n)}|^2 } \left( {\rm Pf}^{\prime} S_n \right)^3 \, .
\eea
Using the expression for $|C_{{\rm GL}(n)}|$ given in (\ref{eq:Jaco-GLn}) and the identity (\ref{eq:X-id}), the formula may be recast into the more compact form
\bea \label{eq:M5-final}
\mathcal{A}^{\rm M5}_{n}
=
 \int d\mu_n^{\rm 6D}  \,\delta^{2 \times n} (V \cdot \Omega \cdot \eta^I ) \, ({\rm Pf}\, U_n)^{-1}\,
 \left( {\rm Pf}^{\prime} S_n \right)^3 \, .
\eea
In this formulation,
\bea
{\cal I}_L^{(2,0)}
= \delta^{2 \times n} (V \cdot \Omega \cdot \eta^I) \, ({\rm Pf}\, U_n)^{-1}\, {\rm Pf}^{\prime} S_n \, ,
\eea
whereas ${\cal I}_R^{\rm DBI}$ remains to be same as that in (\ref{eq:M5-ILIR}). The measure in (\ref{eq:M5-final}) is based on the polarized scattering equations
\bea
\int d\mu_n^{\rm 6D} =\int  {d^n \s d^n v d^{2n} u \over \vol( \slc_{\sigma} \times \slc_{u} )} \delta^{4\times n}
(V \cdot \Omega \cdot \Lambda^A )\, ,
\eea
which is the ${\rm GL}(n, \IC)$ transformation of (\ref{eq:RM-measure}).

Formulas similar to those of the M5-brane can be obtained for tree-level superamplitudes of the D5-brane, which has non-chiral $(1,1)$ supersymmetry. After the ${\rm GL}(n, \IC)$ transformation, the rational-map formula of~\cite{Heydeman:2017yww} reduces to
\begin{align}
\mathcal{A}^{\rm D5}_{n}
&=
 \int d\mu_n^{\rm 6D}  \, \delta^n (V \cdot \Omega \cdot \eta ) \,  \delta^n (\widetilde{V} \cdot
 \widetilde{\Omega} \cdot \tilde{\eta} ) \,
 {V_n \over |C_{{\rm GL}(n)}|  |\widetilde{C}_{{\rm GL}(n)}| }\, \left( {\rm Pf}^{\prime} S_n \right)^3   \cr
&=
 \int d\mu_n^{\rm 6D}  \, \delta^n (V \cdot \Omega \cdot \eta ) \,  \delta^n (\widetilde{V} \cdot \
 \widetilde{\Omega} \cdot \tilde{\eta} ) \, ({\rm Pf}\, U_n)^{-{1\over 2}}\,  ({\rm Pf}\,
 \widetilde{U}_n)^{-{1\over 2}}\,   \left( {\rm Pf}^{\prime} S_n \right)^3  \cr
&=
  \int d\mu_n^{\rm 6D}  \, {\cal I}_L^{(1,1)} \,  {\cal I}_R^{\rm DBI} \, ,
\end{align}
where now
\bea
{\cal I}_L^{(1,1)}
= \delta^{n} (V \cdot \Omega \cdot \eta ) \, ({\rm Pf}\, U_n)^{-1/2}\,
  \delta^{n} (\widetilde{V} \cdot \widetilde{\Omega} \cdot \tilde\eta ) \, ({\rm Pf}\, \widetilde{U}_n)^{-1/2}\, {\rm Pf}^{\prime} S_n\, .
\eea
In the second line we have applied the relation (\ref{eq:Jaco-GLn}) and the identity (\ref{eq:X-id}) as
well as the conjugate version of them. As we have emphasized, ${\cal I}_L^{(1,1)}$ contains conjugate
variables, such as $\tilde{u}_{i, \hat{a}}, \tilde{v}_i$, that appear in $\widetilde{V},  \widetilde{U}_n$. It should be understood that these conjugate variables are determined in terms of the external kinematics $\tilde{\Lambda}_{A}$ via the conjugate
version of the scattering equations (\ref{eq:GLn-general-2}). As made clear in the formula, we do not integrate over these conjugate variables, and the conjugate scattering equations do not appear explicitly either.  The same remark applies to the other non-chiral theories.

Again, the right-hand integrand ${\cal I}_R^{\rm DBI}$ implies that only amplitudes with even number of particles
are non-trivial. Therefore, the original formulas for the M5-brane and D5-brane amplitudes based on rational
maps do not suffer from the issue of an artificial distinction between amplitudes with an even and an odd
numbers of particles. However, this is not the case for the scattering amplitudes of 6D SYM and
supergravity, which we consider next.

\subsection{Maximal SYM}

Let us begin with the 6D $(1,1)$ SYM. The on-shell spectrum of the theory can be packaged in the following form,
\bea \label{eq:6DSYM}
\Phi(\eta, \bar{\eta})= \phi^{1\hat{1}} + \eta_a \psi^{a \hat{1}}+ \tilde{\eta}_{\hat a} \psi^{1 \hat {a} }
+ \eta_a \tilde{\eta}_{\hat a}  A^{a \hat{a}} + \ldots
+ ( \eta)^2(\tilde{\eta})^2 \phi^{2\hat{2}} \, ,
\eea
where, for instance, $A^{a \hat{a}}$ is the 6D gluon.
The theory has non-trivial scattering amplitudes for both even and odd $n$. Beginning with the case of even
$n$, the superamplitude is given by~\cite{Cachazo:2018hqa},
\begin{align} \label{eq:SYM-RM}
\mathcal{A}^{\rm SYM}_{n}  (\alpha)
&=
 \int  d\mu^{\rm 6D}_n  \, \delta^n (C \cdot \Omega \cdot \eta)
 \delta^n (\widetilde{C} \cdot \widetilde{\Omega} \cdot \tilde{\eta})
\,  V_n \,    {\rm Pf}^{\prime} S_n \, {\rm PT} (\alpha)  \, ,
\end{align}
where ${\rm PT} (\alpha)$ is the Parke--Taylor factor, which encodes the color structure of Yang--Mills
amplitudes. Here $\alpha$ represents a permutation of the external particles $\{1,2, \ldots, n\}$. For
instance, when $\alpha$ is the identity permutation,
\bea
{\rm PT} (1,2, \ldots, n) = {1\over \sigma_{12} \sigma_{23} \cdots \sigma_{n-1\, n} \sigma_{n1}} \, .
\eea
The Parke--Taylor factor has conformal weight $w=2$.

Again, the ${\rm GL}(n, \IC)$ transformation leads to
\begin{align} \label{eq:SYM-PS}
\mathcal{A}^{\rm SYM}_{n}  (\alpha)
=
\int d\mu_{n}^{\rm 6D}\, \delta^n (V \cdot \Omega \cdot \eta) \delta^n (\widetilde{V} \cdot
\widetilde{\Omega} \cdot \tilde{\eta})  \,  {V_n  \over |C_{{\rm GL}(n)}| |\widetilde{C}_{{\rm GL}(n)}| } \,
{\rm Pf}^{\prime} S_n \, {\rm PT} (\alpha) \, .
\end{align}
The integrand of the preceding expression can be simplified further using the identity
\begin{align} \label{eq:id-H}
{V_n  \over |C_{{\rm GL}(n)}| |\widetilde{C}_{{\rm GL}(n)}| } \,  {\rm Pf}^{\prime} S_n
=  {|X_{Y, \overline{Y}}| \over |U_{Y, \overline{Y}}| |\widetilde{U}_{Y, \overline{Y}}| }  \,
{\rm Pf}^{\prime} S_n  ={\rm det}^{\prime} H_n \, ,
\end{align}
where the second equality in the above equation is identical to equation $(30)$ of~\cite{Geyer:2018xgb}.
Here $H_n$ is an $n\times n$ matrix introduced by Geyer and Mason in~\cite{Geyer:2018xgb}, which generalizes
the Hodges' determinant of 4D theories~\cite{Hodges:2012ym}. The $H_n$ matrix has the following entries
\bea
H_{ij} = { \langle \epsilon_i  \lambda^A_i \rangle [ \tilde{\epsilon}_j  \tilde{\lambda}_{A,j}  ] \over
\sigma_{ij} } \quad {\rm for}\quad  i \neq j \, , \qquad
u_{i, a} H_{ii} = - {{\lambda}^A_{a}(\sigma_i) [ \tilde{\epsilon}_i  \tilde{\lambda}_{A,i}  ]  } \, .
\eea
Here, just as $\epsilon_{i,a}$, we can choose $\tilde{\epsilon}_{i, \hat{a}} = (0, 1)$. Note that $H_{ii}$ is
independent of the choice of little-group index $a$, namely it is a Lorentz scalar. The reduced determinant
${\rm det}^{\prime}H$ is defined as
\bea
{\rm det}^{\prime} H = { {\rm det} H^{[ij]}_{[kl]}  \over \langle u_i u_j\rangle [\tilde{u}_k \tilde{u}_l]}
\, ,
\eea
where $H^{[ij]}_{[kl]}$ means that we remove the $i$-th and $j$-th columns as well as the $k$-th and $l$-th
rows, and the result is independent of the choices~\cite{Geyer:2018xgb}. Alternatively, using (\ref{eq:id-H}),
when $n$ is even, this can be recast in the more appealing form
\bea \label{eq:Heven}
{\rm det}' H = ({\rm Pf}\, U_n)^{-1/2}\, ({\rm Pf}\, \widetilde{U}_n)^{-1/2}\, {\rm Pf}^{\prime} S_n,
\eea
which has conformal weight $w=2$.

With the help of the identity (\ref{eq:id-H}), the ${\rm GL}(n, \IC)$ transformation leads to the following new formula for the tree superamplitudes of 6D $(1,1)$ SYM,
\begin{align} \label{eq:even-YM}
\mathcal{A}^{\rm SYM}_{n}  (\alpha)
=
\int d\mu^{\rm 6D}_n \, {\cal I}_L^{(1,1)}   \, {\cal I}_R^{(\alpha)} \, ,
\end{align}
where the left and right integrands are given by
\bea
{\cal I}_L^{(1,1)} = \delta^n (V \cdot \Omega \cdot \eta) \delta^n (\widetilde{V} \cdot \widetilde{\Omega} \cdot
\tilde{\eta})\,  {\rm det}^{\prime} H_n \, , \qquad {\cal I}_R^{(\alpha)} = {\rm PT} (\alpha) \, .
\eea
Although the formula (\ref{eq:even-YM}) has been obtained here for even $n$, we will now show that it holds
also for odd $n$!

All that is required is to extend the definition of ${\rm det}' H_n $ to odd $n$,
since all the other factors are already well-defined for odd $n$. Specifically, for odd $n$ we need to replace
(\ref{eq:id-H}) or (\ref{eq:Heven}) by
\bea \label{eq:Hodd}
{\rm det}' H_n = {V_n  \over |C_{{\rm GL}(n)}| |\widetilde{C}_{{\rm GL}(n)}| } \,   {\rm Pf}^{\prime} \widehat{S}_n \, ,
\eea
where $\widehat{S}_n$ is a well-defined expression for odd $n$~\cite{Cachazo:2018hqa}.
This is achieved by extending the $n\times n$ matrix $S_n$ is to an $(n{+}1)\times (n{+}1)$ matrix, which we denote $\widehat{S}_n$. $\widehat{S}_n$ is defined in the same way as ${S}_n$ for even $n$,
as given in (\ref{eq:sij}), but with particle labels $i,j=1,2, \ldots,n$ and $\star$. Here $\sigma_\star$
is a reference puncture, and $p_{\star}$ is a null momentum defined by
\be \label{eq:pstar}
p_{\star}^{AB} = \dfrac{ 2\, q^{[A} p^{B]C}(\sigma_\star) \tilde{q}_C }{ q^D  [\tilde{\rho}_D(\sigma_\star )
\, \tilde{\xi}] \langle \rho^E(\sigma_\star)\, \xi \rangle  \tilde{q}_E}\, ,
\ee
where the bracket $[A\, B]$ means anti-symmetrization, $p^{BC}(\sigma_\star) := \langle \rho^B(\sigma_\star) \rho^C(\sigma_\star)  \rangle$, and $q$ and $\tilde{q}$ are arbitrary spinors.

The validity of these formulas for odd $n$ depends on the identity (\ref{eq:Hodd}). Using the definition of ${\rm Pf}' \widehat{S}_n$ and the odd-point Jacobian given in (\ref{eq:CGL-odd}), one can see that the right-hand side of (\ref{eq:Hodd}) has weight $2$, which matches with that of ${\rm det}' H_n$.  This identity has been verified numerically to hold on the support of the scattering equations. Furthermore, since all the objects in both sides of (\ref{eq:Hodd}) are known to factorize properly, it is straightforward to see that the equality holds in the factorization limits.  Therefore, as we remarked previously, under the ${\rm GL}(n, \IC)$ transformation the rational-map formulas also reduce to (\ref{eq:even-YM}) for odd $n$. This is not so surprising, since all of the ingredients in (\ref{eq:even-YM}), especially $H_n$, make no distinction between even and odd $n$.

Given the formula for superamplitudes in 6D $(1,1)$ SYM, it is straightforward to obtain the superamplitudes for SYM theories in lower dimensions~\cite{Cachazo:2018hqa}, in particular the massive amplitudes in 4D $\mathcal{N}=4$ SYM on the Coulomb branch. The massive spectrum of $\mathcal{N}=4$ SYM on the Coulomb branch can be obtained upon a dimensional reduction of (\ref{eq:6DSYM}),
\bea
\Phi(\eta) = \phi + \eta_a^I \psi_I^a + \eta^I_a \eta^{J, a} \phi_{IJ} +  \eta_{I, a} \eta^{J}_b A^{ab} + \ldots + (\eta)^4 \bar{\phi}\, ,
\eea
here we identify $\{\eta_a, \tilde{\eta}_{\hat a}\}:= \eta^I_a$ with $I=1,2$, and $a=1,2$ is the $SU(2)$ little-group index of a massive particle in 4D.  The superamplitude is then given by
\begin{align} \label{eq:YM-CB}
\mathcal{A}^{\rm {\cal N}=4\, CB}_{n}  (\alpha)
= \int  d\mu^{\rm 4D\,CB}_n  \delta^{2 \times n} (V \cdot \Omega \cdot \eta^I)  \,  {\rm det}^{\prime} H^{\rm CB}_n \,  {\rm PT} (\alpha) \, .
\end{align}
The measure $d\mu^{\rm 4D\,CB}_n$ takes a form similar to that of the 6D measure of massless particles,
\bea
\int d \mu^{\rm 4D\,CB}_{n} = \int  {  d^n \s \, d^{n} v \, d^{2n} u
\over {\rm vol} ( \slc)_\s \times  \slc_u  ) } \delta^{2 \times n} ( V \cdot \Omega \cdot  \Lambda^{\alpha})\delta^{2 \times n} ( V \cdot \Omega \cdot  \widetilde{\Lambda}^{\dot \alpha}) \, ,
\eea
where $ \Lambda^{\alpha}$ and $\widetilde{\Lambda}^{\dot \alpha}$ are the reduction of the 6D massless kinematics $\Lambda^A$ via
\bea
\lambda^A_a \rightarrow \begin{pmatrix}
\lambda_{\alpha, 1} & \lambda_{\alpha, 2} \\
\tilde{\lambda}^{\dot \alpha}_1 & \tilde{\lambda}^{\dot \alpha}_2
\end{pmatrix} \, ,
\eea
for each particle.\footnote{For $\tilde{\lambda}_{A, \hat a}$, we perform an analogous reduction $\tilde{\lambda}_{A, \hat a} \rightarrow \begin{pmatrix}
\lambda^{\alpha}_1 & \lambda^{\alpha}_2 \\
\tilde{\lambda}_{\dot \alpha, 1} & \tilde{\lambda}_{\dot \alpha, 2}
\end{pmatrix}$. }
The 4D momentum and mass are given by
\bea
p_{i}^{\alpha \dot \alpha} = \l^{\alpha}_{i, a} \tilde{\l}^{\dot \alpha}_{i, b} \epsilon^{ab}\, , \quad 
M_{i} \epsilon^{\alpha \beta} = \l^{\alpha}_{i, a} {\l}^{ \beta}_{i, b} \epsilon^{ab} \, , \quad
\tilde{M}_{i} \epsilon^{\dot \alpha \dot \beta} = \tilde{\l}^{\dot \alpha}_{i, a} \tilde{\l}^{\dot \beta}_{i, b} \epsilon^{ab} \, ,
\eea
and $M_{i} \tilde{M}_{i} = m_i^2$ is the mass square. The masses arise as the extra dimensional momenta satisfying  $\sum_i M_{i} = \sum_i \tilde{M}_{i}=0$. 
Here we have decomposed the 6D spinor indices according to $A=\{\alpha, \dot \alpha\}$. Also we identified $a$ and $\hat{a}$, since they both reduce to the same little-group index of a 4D massive particle. For massless particles we further set $\lambda_{\alpha, 1} = \tilde{\lambda}_{\dot \alpha, 2}=0$. The integrand $H^{\rm CB}$ is obtained from $H$, again via dimensional reduction,
\bea
H^{\rm CB}_{ij} = { \epsilon_{i, a} \epsilon_{j, b}  ( \langle i^a j^b \rangle+ [ i^a j^b] )   \over
\sigma_{ij} } \quad {\rm for}\quad  i \neq j \, , \, \, \,
u_{i, a} H^{\rm CB}_{ii} = - \epsilon_{i, b} (\langle {\lambda}_{a}(\sigma_i) \,  i^b \rangle + [\tilde{\lambda}_{a}(\sigma_i) \,  i^b]) \, ,
\eea
where $\langle i^a j^b \rangle := \lambda^{ a}_{\alpha, i} \lambda^{\alpha, b}_{j}$ and  $[ i^a j^b ]:= \tilde{\lambda}^{ a}_{\dot \alpha, i} \tilde{\lambda}^{\dot \alpha, b}_{j}$. The  reduced determinant is defined as
\bea
{\rm det}^{\prime} H^{\rm CB} = { {\rm det}\, (H^{\rm CB})^{[ij]}_{[kl]}  \over \langle u_i u_j\rangle  \langle u_k u_l\rangle }
\, .
\eea
Due to the dimensional reduction, $\tilde{u}_i$ has been identified with ${u}_i$.

\subsection{Maximal supergravity}

The construction of a concise formula for the tree amplitudes of 6D maximal supergravity, namely $(2,2)$ supergravity, is now very easy. It can be obtained from the formula for 6D $(1,1)$ SYM via the standard
double-copy procedure in the CHY formalism of scattering amplitudes. The procedure is to replace,
${\cal I}_R^{(\alpha)}$, the Parke--Taylor factor ${\rm PT}(\alpha)$ by ${\cal I}_R^{(1,1)}$,
\bea
{\rm PT}(\alpha) \rightarrow \delta^n (V \cdot \Omega \cdot \eta) \delta^n (\widetilde{V} \cdot
\widetilde{\Omega} \cdot \tilde{\eta}) \, {\rm det}^{\prime} {H}_n \, .
\eea
This leads to a formula for the $(2,2)$ supergravity tree-level superamplitudes that is valid for all $n$,
\begin{align} \label{eq:22SUGRA}
\mathcal{M}^{(2,2)}_{n}
&=\int d\mu_n^{\rm 6D}  \delta^{2 \times n} (V \cdot \Omega \cdot \eta^I) \delta^{2 \times n}
(\widetilde{V} \cdot  \widetilde{\Omega}  \cdot \tilde{\eta}^{\tilde I})  \,
({\rm det}^{\prime} H_n)^2 \, ,
\end{align}
where $I=1,2$ and $\tilde{I}=\tilde{1}, \tilde{2}$. This formula for the $(2,2)$ supergravity amplitudes will be used for constructing the superamplitudes for $(2,0)$ supergravity, as described in the next subsection.

\subsection{6D (2,0) Supergravity}

The 6D rational-map formalism has also been applied to non-maximal supersymmetric theories, in particular
the 6D $(2,0)$ supergravity~\cite{Heydeman:2018dje}. It arises as the low-energy theory of type IIB
superstring theory compactified on a $K3$ surface, with interactions between graviton
supermultiplets and $21$ flavors of tensor supermultiplets. The superfield of the tensor multiplet, given in
(\ref{eq:M5-field}), is a scalar, whereas the graviton supermultiplet transforms as a $({\bf 1}, {\bf 3})$
representation of the little group. Therefore it carries explicit ${\rm SU}(2)_R$ little-group
indices,
\bea
\Phi_{\hat{a} \hat{b}}(\eta)
=  B_{\hat{a} \hat{b}} + \ldots+ \eta^{I}_a \eta_{I,b} G^{ab}_{\hat{a} \hat{b}} + \ldots  + (\eta)^4
\bar{B}_{\hat{a} \hat{b}} \, .
\eea
In particular, the bottom component $B_{\hat{a} \hat{b}}$ corresponds to an anti self-dual two-form and $G^{ab}_{\hat{a} \hat{b}}$ in the middle, which transforms as $({\bf 3}, {\bf 3})$, is the 6D graviton.

As shown in~\cite{Heydeman:2018dje}, the spectrum and the amplitudes of $(2,0)$ supergravity interacting
with a single tensor multiplet can be obtained from those of $(2,2)$ supergravity via a SUSY reduction,
\bea \label{eq:fermionic}
\mathcal{M}^{(2,0)}_{n_1, n_2} =\int \prod_{i \in n_1} d\tilde{\eta}^{\hat{I}}_{i, \hat{a}_i }
d \tilde{\eta}_{i, {\hat{I}} \hat{b}_i} \prod_{j \in n_2} d\tilde{\eta}^{\hat{1}}_{j, \hat{a}_j }
d \tilde{\eta}^{ \hat{2} \, \hat{a}_j}_{j}    \mathcal{M}^{(2,2)}_n \, .
\eea
Here $n_1$ is the number of graviton multiplets and $n_2$ is the number of tensor multiplets, with
$n_1+n_2=n$. From the explicit formula of $\mathcal{M}^{(2,2)}_n$ given in (\ref{eq:22SUGRA}), the above
fermionic integration leads to
\begin{align} \label{eq:(20)-single}
\mathcal{M}^{(2,0)}_{n_1, n_2}  &=  \int d \mu_n^{\rm 6D} \, \widetilde{V}^{n_1, n_2}_{\hat{a}, \hat{b}} \,
 \delta^{2 \times n} (V \cdot \Omega \cdot \eta^I)   \,
({\rm det}^{\prime} H_n )^2 \, .
\end{align}
Here $\widetilde{V}^{n_1, n_2}_{\hat{a}, \hat{b}}$ arising from the fermionic integration is given by
\bea \label{eq:V-contraction}
\widetilde{V}^{n_1, n_2}_{\hat{a}, \hat{b}} = |\widetilde{V}_{\hat{a}_1 \cdots \hat{a}_n}| \,
 |\widetilde{V}_{\hat{b}_1 \cdots \hat{b}_n}|  \prod_{i \in n_2} \epsilon^{\hat{a}_i \hat{b}_i}
 + {\rm sym}\, ,
\eea
where the matrix $\widetilde{V}_{\hat{a}_1 \hat{a}_2, \ldots,  \hat{a}_n}$ is defined as
\bea
 \widetilde{V}_{\hat{a}_1 \cdots \hat{a}_n} = \left( \begin{array}{cccc}
\widetilde{V}_{ 1;1, \hat a_1}  &\widetilde{V}_{1;2, \hat a_2} & \cdots & \widetilde{V}_{1;n, \hat a_n} \\
\widetilde{V}_{ 2;1, \hat a_1}  &\widetilde{V}_{2;2, \hat a_2} & \cdots & \widetilde{V}_{2;n, \hat a_n} \\
\vdots  & \vdots  & \cdots & \vdots  \\
\widetilde{V}_{ n;1, \hat a_1}  &\widetilde{V}_{n;2, \hat a_2} & \cdots & \widetilde{V}_{n;n, \hat a_n}
\end{array}  \right) \, .
\eea
In (\ref{eq:V-contraction}), we contract the little-group indices $\hat{a}_i, \hat{b}_i$ for $i \in \{1,2,\dots,n_2\}$, because they refer to tensor multiplets, whose superfield is a scalar. The symbol ``${\rm sym}$" means that we symmetrize the little-group indices $\hat{a}_j, \hat{b}_j$ with $j \in \{1,2,\dots,n_1\}$ of the graviton multiplets. Therefore,
the subscripts $\hat{a}, \hat{b}$ of $V^{n_1, n_2}_{\hat{a}, \hat{b}}$ are uncontracted
little-group indices of graviton multiplets.

As pointed out in~\cite{Heydeman:2018dje}, when we formulate the tree amplitudes of $(2,0)$ supergravity
using rational maps, the object ${\rm Pf} {X}_{n_2}$ arises after the fermionic integration
(\ref{eq:fermionic}). Here ${X}_{n_2}$ is an antisymmetric $n_2 \times n_2$ matrix and is defined as
\bea
\left[X_{n_2}\right]_{ij} =\begin{cases}
               {1 \over \sigma_{ij} } \quad ~  {\rm if}  \quad i\neq j \, , \\
            0  \quad \quad {\rm if}  \quad i = j\, .
            \end{cases}
\eea
The importance of ${\rm Pf} {X}_{n_2}$ is that it allows us to introduce multiple flavors for the tensor
multiplets of $(2,0)$ supergravity via a replacement $X_{n_2} \rightarrow \mathcal{X}_{n_2}$, with
$\mathcal{X}_{n_2}$ given by
\bea
\left[\mathcal{X}_{n_2}\right]_{ij} =\begin{cases}
               {\delta_{f_i f_j} \over \sigma_{ij} } \quad ~  {\rm if}  \quad i\neq j \, , \\
            0  \quad \quad ~~ {\rm if}  \quad i = j\, ,
            \end{cases}
\eea
where $f_i, f_j$ are flavor indices of particles $i, j$. $\delta_{f_i f_j}=1$ if the particles have the
same flavor and otherwise  $\delta_{f_i f_j}=0$. This procedure was also applied to scattering amplitudes
of Einstein--Maxwell theory with multiple $U(1)$ photons~\cite{Cachazo:2014xea}.

We see that the net effect of the above procedure for going from a theory with a single tensor multiplet to
a theory with multiple tensor multiplets is to insert the factor
\bea
{ {\rm Pf} \mathcal{X}_{n_2} \over {\rm Pf} X_{n_2}} ,
\eea
where the subscript $n_2$, which is required to be even, represents the number of the particles belonging to tensor multiplets. Therefore, by this procedure and the single-flavor formula (\ref{eq:(20)-single}), we obtain a formula for the tree
amplitudes of 6D $(2,0)$ supergravity describing the interaction of graviton multiplets with multi-flavor
tensor multiplets,
\begin{align}
\mathcal{M}^{(2,0)}_{n_1, n_2}  &=  \int d \mu_n^{\rm 6D} \,{ {\rm Pf} \mathcal{X}_{n_2} \over {\rm Pf}
X_{n_2}}\, V^{n_1, n_2}_{\hat{a}, \hat{b}} \,  \delta^{2 \times n} (V \cdot \Omega \cdot \eta^I)   \,
({\rm det}^{\prime} H_n)^2 \, .
\end{align}
We have checked numerically that this new formula gives the same results as obtained from the formula
proposed in~\cite{Heydeman:2018dje}, namely equation (26) of that paper. The result may also be expressed in a double copy form,\footnote{The BCJ double copy structure~\cite{Bern:2008qj} of 6D $(2,0)$ supergravity was also studied in~\cite{Johansson:2017bfl}}
\begin{align} \label{eq:(2,0)-LR}
\mathcal{M}^{(2,0)}_{n_1, n_2}  &=  \int d \mu_n^{\rm 6D} \, {\cal I}^{(1,0), f}_{L, \hat a} \, {\cal I}^{(1,0)}_{R, \hat b}  \prod_{i \in n_2} \epsilon^{\hat{a}_i \hat{b}_i}
 + {\rm sym} \, ,
\end{align}
where each half integrand has $(1,0)$ supersymmetry, and one of them (${\cal I}^{(1,0), f}_{L, \hat a}$) carries the flavor symmetry,
\begin{align}
{\cal I}^{(1,0), f}_{L, \hat a} = \delta^{n} (V \cdot \Omega \cdot \eta)   \,
{\rm det}^{\prime} H_n |\widetilde{V}_{\hat{a}_1 \cdots \hat{a}_n}| { {\rm Pf} \mathcal{X}_{n_2} \over {\rm Pf}
X_{n_2}} \, , \quad
{\cal I}^{(1,0)}_{R, \hat b} = \delta^{n} (V \cdot \Omega \cdot \eta)   \,
{\rm det}^{\prime} H_n |\widetilde{V}_{\hat{b}_1 \cdots \hat{b}_n}| \, .
\end{align}
Again in (\ref{eq:(2,0)-LR}) we symmetrize the little group indices for graviton multiplets and for tensor multiplets we contract the indices.

\section{Conclusion} \label{sec:conclusion}

The Grassmannian formulation of scattering amplitudes of 4D $\mathcal{N}=4$ SYM was proposed and extensively
studied in~\cite{ArkaniHamed:2009dn, ArkaniHamed:2012nw}. It has led to powerful geometric pictures for the
scattering amplitudes of that theory. Grassmannian formulations were later extended to amplitudes of 3D
supersymmetric Chern--Simons matter theories~\cite{Lee:2010du, Huang:2013owa, Huang:2014xza}. In this
formulation and its on-shell diagram realization, scattering amplitudes are described in terms of on-shell
data only, and physical concepts such as unitarity and locality emerge as derived consequences of geometric
properties of the Grassmannian. The Grassmannian formulation of 4D $\mathcal{N}=4$ SYM was further generalized, leading to a new
mathematical object for scattering amplitudes, called the amplituhedron, in~\cite{Arkani-Hamed:2013jha}.

It is natural to ask whether these ideas extend to other theories, especially 6D $(1,1)$ SYM and 4D $\mathcal{N}=4$ SYM on the Coulomb branch. Previously, we have seen the appearance of the symplectic Grassmannian in the description of the scattering amplitudes of these theories~\cite{Cachazo:2018hqa}. In this paper, we showed that the understanding in terms of the symplectic Grassmannian is fruitful and provides a unifying picture for two different-looking twistor formulations. One of them is based on rational maps and the other is based on polarized scattering equations. In particular, these two distinct formulations simply reflect different choices for fixing the ${\rm GL}(n, \IC)$ symmetry in our construction of the symplectic Grassmannian $\mathbb{LG}(n, 2n)$. This understanding was shown to lead to new formulas for 6D superamplitudes of other interesting theories that were originally studied in~\cite{Heydeman:2017yww, Cachazo:2018hqa, Heydeman:2018dje} including world-volume
DBI-like theories of single probe branes, $(1,1)$ SYM,  $(2,2)$ and $(2,0)$ supergravity. Upon dimensional reduction, following the procedure of \cite{Cachazo:2018hqa}, we also obtained formulas for scattering amplitudes of 4D $\mathcal{N}=4$ SYM on the Coulomb branch. The spectrum includes both massive and massless supermultiplets.

This paper has focused on the implications of the symplectic Grassmannian for the scattering equations that enter in the integrands of the scattering amplitudes, despite the fact that the integrands are still expressed in terms of world-sheet variables. It would be interesting to investigate how to recast the integrands in terms of Pl${\rm \ddot{u}}$cker coordinates of the symplectic Grassmannian. This should lead to a Grassmannian representation of 6D superamplitudes with local ${\rm GL}(n, \IC)$ symmetry. As in the case of 4D $\mathcal{N}=4$ SYM and 3D supersymmetric Chern--Simons matter theory, such a symplectic Grassmannian formulation should be closely related to on-shell diagram representations of BCFW recursion relations~\cite{Britto:2005fq}, especially the 6D version of the recursion relations~\cite{Cheung:2009dc, Dennen:2009vk}. Furthermore, using the symmetric plabic graphs that were introduced in~\cite{Karpman:2015oha}, the stratification of positive symplectic Grassmannian has been explored in~\cite{stratificationLG}, which may be directly related to the on-shell diagram representation of 6D amplitudes. 

It would also be interesting to extend the ideas developed in this paper to the polarized scattering equations in ten and eleven dimensions, which were proposed in~\cite{Geyer:2019ayz} (see~\cite{Mizera:2019gea} for a different approach to scattering amplitudes in higher dimensions). This should help to develop a better understanding of the structure of scattering amplitudes in higher-dimensional supersymmetric theories as well as their reformulations in terms of geometric objects such as Grassmannians.

Finally, it has been found recently that the 6D tree-level supergravity amplitudes have important implications for the correlation functions in $AdS_3 \times S^3$~\cite{Rastelli:2019gtj, Giusto:2019pxc} (see~\cite{Rastelli:2016nze, Caron-Huot:2018kta} for applications of tree-level amplitudes in 10D
supergravity to the correlators in $AdS_5 \times S^5$). It would be very interesting if any of the new structures of 6D superamplitudes discussed in this paper could be applied to holographic correlators in $AdS_3 \times S^3$.

\section*{Acknowledgements}

We would like to thank Freddy Cachazo, Alfredo Guevara, Matthew Heydeman, and Sebastian Mizera for collaboration at an early stage of this project. We also thank Henrik Johansson, Lionel Mason and Shun-Qing Zhang for helpful discussions. We are grateful to Freddy Cachazo, Alfredo Guevara, and Sebastian Mizera for insightful comments on early drafts. C.W. is supported by a Royal Society University Research Fellowship No. UF160350. J.H.S. is supported in part by the Walter Burke Institute for Theoretical Physics at Caltech and by U.S. DOE Grant DE-SC0011632. We thank the Galileo Galilei Institute for Theoretical Physics and the INFN for hospitality and partial support during the workshop ``String theory from a worldsheet perspective," where part of this work has been done.

\end{document}